# MRO/CRISM Retrieval of Surface Lambert Albedos for Multispectral Mapping of Mars with DISORT-based Radiative Transfer Modeling: Phase 1 – Using Historical Climatology for Temperatures, Aerosol Optical Depths, and Atmospheric Pressures

April 30, 2008

**Patrick C. McGuire,**
*McDonnell Center for the Space Sciences, Washington University in St. Louis*
**Michael J. Wolff,**
*Space Science Institute*
**Michael D. Smith,**
*NASA/Goddard Space Flight Center*
**Raymond E. Arvidson,**
*McDonnell Center for the Space Sciences, Washington University in St. Louis*
**Scott L. Murchie,**
*Applied Physics Lab, John Hopkins University*
**R. Todd Clancy,**
*Space Science Institute*
**Ted L. Roush,**
*NASA Ames Research Center*
**Selby C. Cull, Kim A. Lichtenberg, Sandra M. Wiseman,**
*McDonnell Center for the Space Sciences, Washington University in St. Louis*
**Robert O. Green, Terry Z. Martin, Ralph E. Milliken[1],**
*NASA/Jet Propulsion Laboratory, California Institute of Technology*
**Peter J. Cavender, David C. Humm, Frank P. Seelos,
Kim D. Seelos, Howard W. Taylor,**
*Applied Physics Lab, John Hopkins University*
**Bethany L. Ehlmann, John F. Mustard, Shannon M. Pelkey[2],**
*Dept. of Geological Sciences, Brown University*
**Timothy N. Titus,**
*U.S. Geological Survey, Flagstaff*
**Christopher D. Hash, Erick R. Malaret,**
*Applied Coherent Technology Corporation*
**and the CRISM Team.**

---

[1] Formerly at Brown University.
[2] Now at GeoEye Inc., 12076 Grant St. Thornton, CO 80241.

**Abstract:**
We discuss the DISORT[3]-based radiative transfer pipeline ('CRISM_LambertAlb') for atmospheric and thermal correction of MRO/CRISM data acquired in multispectral mapping mode (~200 m/pixel, 72 spectral channels). Currently, in this phase-one version of the system, we use aerosol optical depths, surface temperatures, and lower-atmospheric temperatures, all from climatology derived from Mars Global Surveyor Thermal Emission Spectrometer (MGS-TES) data, and surface altimetry derived from MGS Mars Orbiter Laser Altimeter (MOLA). The DISORT-based model takes as input the dust and ice aerosol optical depths (scaled to the CRISM wavelength range), the surface pressures (computed from MOLA altimetry, MGS-TES lower-atmospheric thermometry, and Viking-based pressure climatology), the surface temperatures, the reconstructed instrumental photometric angles, and the measured *I/F* spectrum, and then outputs a Lambertian albedo spectrum. The Lambertian albedo spectrum is valuable geologically since it allows the mineralogical composition to be estimated. Here, *I/F* is defined as the ratio of the radiance measured by CRISM to the solar irradiance at Mars divided by $\pi$; if there was no martian atmosphere, *I/F* divided by the cosine of the incidence angle would be equal to the Lambert albedo for a Lambertian surface. After discussing the capabilities and limitations of the pipeline software system, we demonstrate its application on several multispectral data cubes – particularly, the outer reaches of the northern ice cap of Mars, the Tyrrhena Terra area northeast of the Hellas basin, and an area near the landing site for the Phoenix mission in the northern plains. For the icy spectra near the northern polar cap, aerosols need to be included in order to properly correct for the $CO_2$ absorption in the $H_2O$ ice bands at wavelengths near 2.0 μm. In future phases of software development, we intend to use CRISM data directly in order to retrieve the spatiotemporal maps of aerosol optical depths, surface pressure and surface temperature. This will allow a second level of refinement in the atmospheric and thermal correction of CRISM multispectral data.

---

[3] For a list of acronyms, see the appendix.

## *1. Introduction*

The Mars Exploration Rovers, Spirit and Opportunity, have been blazing two different trails on the martian surface for almost two Mars years now. Spirit has found that the Gusev Crater does not contain sedimentary deposits from Ma'adim Valles as initially believed, but rather that basaltic materials cover the crater floor (*Squyres et al.*, 2004a) and the Columbia Hill outcrops exhibit a variety of interesting water-related alteration minerals (*Squyres et al.*, 2006). Opportunity has confirmed the orbital-based detection of coarse-grained gray hematite at Terra Meridiani (*Squyres et al.*, 2004b), and it has explored the depths of several small to medium-sized craters, allowing extensive stratigraphic analyses of the sediments (*Squyres et al.*, 2004b; *Grotzinger et al.*, 2005). The instrumentation on both rovers has allowed truly spectacular mineralogical studies to be completed and the geological understanding of two sites on Mars to be markedly improved, thus giving ground truth for several orbital missions' studies of the martian atmosphere, surface, and subsurface (e.g., *Christensen et al.*, 2003; *Bibring et al.*, 2005; *M. Smith et al.*, 2001; *M. Smith,* 2004).

These two rovers have blazed trails for future ground-based missions as well. The experience gained from these missions, both from an engineering perspective and from a scientific perspective, will offer greater probabilities of success for the recently-launched Phoenix Lander 2007, for the upcoming Mars Science Laboratory 2009, and for Exomars 2013. However, from a geological and mineralogical perspective, the impact of the current and past orbital missions may do more to assist the upcoming landed missions than the spatially limited observations from either Spirit or Opportunity. From the bird's-eye perspective, orbital scientific instruments, such as HiRISE, HRSC, MGS-TES, THEMIS, MOLA, MOC, PFS, GRS, MARSIS, SHARAD, OMEGA, and CRISM, have the capability to characterize and evaluate future landing sites. These orbital instruments have global mapping capabilities of sufficient resolution to have been critical in the selection of a safe and scientifically interesting landing site for the ice-prospecting operations of the Phoenix Lander 2007 (*P. Smith, 2003)*. Similar geomorphic and mineralogical studies of data acquired with instrumentation in Mars orbit are currently providing a wealth of interesting landing-site options for the Mars Science Laboratory 2009.

The CRISM imaging spectrometer has been acquiring data from its platform on the Mars Reconnaissance Orbiter (MRO) since September 2006 (*Murchie et al.*, 2007). The data consists of both: (i) hyperspectral targeted observations with high spatial resolution (~18 m/pixel) and 544 spectral channels, and (ii) multispectral mapping strips with lower spatial resolution (~200 m/pixel) and 72 spectral channels. The spectral coverage for both modes of observation is 0.362-3.920 µm. The data from both observing modes have been extremely valuable to the Mars exploration community in the selection of landing sites for the Mars Science Laboratory (*F. Seelos, Barnouin-Jha, & Murchie*, 2008; *Golombek et al.*, 2008), particularly due to the capability of CRISM to identify minerals (i.e., phyllosilicates, sulfates, etc.) on the surface of Mars. The high spatial resolution of CRISM is particularly useful for the landing site evaluation. It should be stressed however, that CRISM has other capabilities besides landing site selection, including the comprehensive spectral targeting and mapping of the entire planet in the near infrared at significantly higher spatial resolution than before. This high spatial resolution in the near infrared will allow for the detection of phyllosilicates, sulfates and other minerals which contain $H_2O$,[4] among other minerals.

In order to quantitatively understand the data from any instrument, investigators need to account for variable observing conditions (cf. *Abdou et al., 2006; Gao et al., 2007; Grey et al., 2006; Lyzenga, Malinas, & Tanis, 2006; Philpot, 2007; Pitman et al., 2007; Salomon et al., 2006)*. From Mars orbit, an instrument that acquires data at visible and near-infrared wavelengths encounters variable photometric angles (solar incidence angle, viewing angle), atmospheric conditions (surface pressure, dust and ice aerosol optical depths), and thermal conditions (surface temperature, surface slope). In this paper, we will describe the photometric, atmospheric and thermal correction system, as first discussed by *McGuire et al.* (2006), that is being used for converting the calibrated CRISM multispectral mapping *I/F* data (*Malaret et al., 2008; Murchie et al.*, 2007) to a true representation of the surface reflectance spectra. This basic pipeline may provide a useful recipe for similar correction pipelines for visible-to-infrared instrumentation that observes Mars, either from a rover, from Mars orbit, or from the Earth/Moon system. Furthermore, the visible-infrared instruments are common to many missions and this pipeline could be generalized to other solar system targets. After accounting for variable observing conditions, mapping and data-mining of both the atmosphere and surface of Mars (e.g., *Bandfield*, 2007; *Bibring et al.*, 2005; *Brown*, 2006; *Christensen et al.*, 2003; *Clancy et al.*, 2003; *Malaret et al.*, 2008; *M. Smith et al.*, 2001; *M. Smith,* 2004; *Pelkey et al.*, 2007; *Schmidt, Douté, & Schmitt,* 2007) can begin in earnest, without concern for the distorting effects of changing surface geometries and atmosphere[5]. With proper correction, we can overlay or mosaick CRISM mapping image strips from different orbits at different times of the martian year of nearby locations, and the resulting digital map will have much reduced discontinuities between the different image strips. Such an endeavor will enhance our ability to accurately map (*Malaret et al.,* 2008*)* the surface properties of Mars and allow a more robust evaluation of future landing sites.

The *I/F* data from CRISM consist of the measured radiance-on-sensor, *I*, divided by the solar radiance, *F*, which is the solar irradiance (*J*) per steradian: *(F = J/π)*. The *I/F* data from CRISM are measured simultaneously for up to 544 different wavelengths from 0.362-3.92 μm, giving a spectrum in the visible to near-infrared which allows studies of both the minerals in the surface and the constituents in the martian atmosphere (*Murchie et al.*, 2007b; *Mustard et al.*, 2007). After the photometric, atmospheric and thermal correction of CRISM multispectral *I/F* data, the resulting $A_L$ data is the incidence-angle corrected reflectance called 'Lambertian albedo'[6]. Retrieval of spectra of Lambert albedos allows: (i) the estimation of mineralogical abundances, and (ii) the quantitative intercomparison of spectra from different orbits of the spacecraft under different observing conditions. The photometric model for a flat Lambertian surface without an atmosphere is $I/F = A_L \cos(INC)$, where *INC* is the solar incidence angle relative to the surface normal (*Hapke*, 1993, pp. 190-191). This photometric model presumes that the diffusely-reflected photons have equally-probable surface-reflectance angles, called emission angles[7] (EMI). This independence of *I/F* on EMI often occurs for bright, granular surfaces. Other more sophisticated non-Lambertian photometric models have been developed, for example the Lunar-Lambert model (*McEwen et al.*, 1991) or the Hapke model (*Hapke*, 1993), but for our multispectral mapping work with CRISM, we will begin with the simpler Lambertian model, with the intent that we will later extend this work to non-Lambertian models, as warranted. For a surface with an atmosphere overlying it, the correction is somewhat more complex, especially for variable observing conditions. The effects of atmospheric scattering and absorption by both molecules and aerosols over different surfaces have been analyzed previously; for example, see: *Bohren & Clothiaux* (2006), *Chandrasekhar* (1960), *Thomas & Stamnes* (1999).

---

[4] $H_2O$ ice has strong absorptions at 1.5 and 2.0 μm. When the mineral is not ice, but instead, contains $H_2O$ either in a chemically-bound or physically-bound form (i.e., hydrated phyllosilicates & sulfates), then these absorptions at 1.5 and 2.0 μm are modulated and shifted to different wavelengths in the near infrared.

[5] The variable atmospheric conditions which are corrected for here in this paper include varying atmospheric $CO_2$ pressure and varying dust and ice aerosol optical depths. By correcting for these major effects of the atmosphere, we can more readily assess more subtle atmospheric components of water vapor, carbon monoxide, oxygen/ozone, etc.

Further detailed work on radiative transfer modeling through the martian atmosphere and off of surfaces of constant albedo has been done in Figure 1 of *Arvidson et al.* (2006) for a case where the optical depth of dust aerosols is 0.35 and the optical depth for ice aerosols is 0.05, with the aerosol optical depths measured at the MGS-TES reference wavelengths of 9.3 and 12.1 μm, respectively. The modeling work by *Arvidson et al.* (2006) was done in order to quantitatively extract Lambert albedos from data from the OMEGA spectrometer. *Arvidson et al.* (2006) found that for these aerosol optical depths, for 'darker' surfaces of Lambert albedo less than a 'transition Lambert albedo' of $A_{L,trans,1}$ ~ 0.15, the modeled *I/F* is higher than the Lambert albedo by up to 80%, with the greatest effect for the lowest albedos[8]. *Arvidson et al.* (2006) also found that for these aerosol optical depths, for 'brighter' surfaces of Lambert albedo greater than the transition Lambert albedo of $A_{L,trans,1}$ ~ 0.15, the modeled *I/F* is lower than the Lambert albedo by up to 20%, with the greatest effect for the highest albedos. These changes in the *I/F* from the expected values without an atmosphere are caused by the aerosols. These general statements apply for wavelengths between 0.6-2.6 μm exempting the $CO_2$ gas bands between 1.9-2.2 μm. For shorter wavelengths between 0.4-0.6 μm, a similar behavior applies but with different amplitudes and a transition Lambert albedo of $A_{L,trans,0} = 0.08$ instead of the longer-wavelength transition Lambert albedo of $A_{L,trans,1} = 0.15$. For wavelengths in the $CO_2$ gas bands between 1.9-2.2 μm, *Arvidson et al.* (2006) modeled that the absorption can be rather deep, with 50% relative absorption by $CO_2$ for 6surfaces of Lambert albedo equal to 0.35, and a 30% relative absorption by $CO_2$ for surfaces of Lambert albedo equal to 0.05.

---

[6] Hereafter, we will often use the term 'Lambert albedo' instead of 'Lambertian albedo', for short.
[7] also known as 'emergence angles'
[8] The term 'transition Lambert albedo' means the value of the Lambert albedo (for particular levels of the ice and dust aerosols) that divides the brightened 'darker' surfaces from the darkened 'brighter' surfaces. $A_{L,trans,1}$ is defined as the value of this transition Lambert albedo for most wavelengths between 0.6-2.2 μm. $A_{L,trans,0}$ is defined as the value of this transition Lambert albedo for wavelengths between 0.4-0.6 μm.

All in all, this variance of behaviors of relatively darker *I/F* for bright surfaces and relatively brighter *I/F* for dark surfaces (in the presence of aerosols) is a primary reason for the detailed radiative transfer correction of the CRISM data that we present here. Aerosols reduce contrast; the radiative transfer correction of CRISM data restores much of this contrast. Furthermore, the amplitude of darkening over 'brighter' surfaces and the brightening over 'darker' surfaces depends on the aerosol content in the martian atmosphere as well as the photometric viewing angles (*Arvidson et al.* 2006; *Wolff et al.*, 2006). *Wolff et al.* (2006) applied a one-band version of the technique which we describe here for MGS-TES broadband albedos over the non-polar regions of Mars (60S – 60N), and found that the bright areas of Mars have solarband MGS-TES Lambert albedos that are typically 5% brighter than without atmospheric correction (see *Wolff et al.* (2006) Figure A1). *Wolff et al.* (2006) also find that dark areas of Mars have solarband MGS-TES Lambert albedos that are up to 30% darker than without atmospheric correction. Without radiative transfer correction of aerosol effects over surfaces of variable brightness, the quantitative accuracy of CRISM multispectral products would be significantly diminished.

The structure of our paper is as follows. First, we give a general overview of the pipeline processing used for CRISM multispectral mapping to correct for surface pressure, aerosols, thermal emission, and photometric angles. Second, we discuss the DISORT-based radiative transfer retrieval of Lambert albedos. Third, we describe the climatology of pressure, temperature and aerosols, where 'climatology' is defined as the historical record of these quantities at different places on Mars as a function of Mars year, solar longitude (day of year), and time of day. Fourth, we describe some results from the application of the correction system to several different CRISM multispectral image cubes or map tiles, including a particular example of correcting a CRISM map tile for an area near the landing site for the Phoenix Lander 2007. Finally, we summarize our results and discuss future work.

## *2. General overview of pipeline processing used to correct for surface pressure, aerosols, thermal emission and photometric angles.*

In Figure 1, we show the processing flowchart for the photometric, atmospheric, and thermal correction system described in this paper. The multispectral data cube in raw EDR format[9] is calibrated to *I/F* format (preliminary TRDR[10]), which is then converted to a Lambert albedo ($A_L$) TRDR, by correcting for:
1. Photometric effects of different viewing and incidence angles,
2. Atmospheric effects of scattering and absorption by $CO_2$ and aerosols, as well as

---

[9] Experimental Data Record (EDR).
[10] Targeted Reduced Data Record (TRDR).

3. Thermal effects caused by emission of photons by surfaces of different temperature.

The intent of this paper is to describe these different correction modules of our system, and to give some examples from the initial applications of this system for retrieving Lambert albedos to multispectral CRISM data. In later papers, we will summarize more comprehensive results, system performance, and the utilization of this system for further scientific studies than those presented here. The scientific ends of this system include improvements in mapping the mineralogy of the martian surface and climatological mapping of gas abundances in the martian atmosphere (*Malaret et al.,* 2008).

Also in a later paper, we will discuss in more detail the system for temperature estimation that is summarized in Figure 1. This paper focuses on the blue colored lines in Figure 1, which account for the correction for photometric angles, for surface pressure, and for aerosols. Climatological correction for thermal emission is also included as part of the current system, as outlined also by the blue colored lines.

The purple, red and green colored lines in Figure 1 represent how the correction for thermal emission using higher spatial-resolution techniques for temperature estimation will be phased into the CRISM_LambertAlb system. These higher spatial-resolution techniques for temperature estimation are not yet in use, largely since the thermal spectral bands (at wavelengths > 3.0 μm) have just recently been calibrated. Currently, we rely on the simplest choice of temperature, which is to use the ADR CL[11] lookup table to look up the climatological value of MGS-TES surface temperature (*M. Smith et al.*, 2004). This is represented by a portion of the blue lines in Figure 1.

We do not correct the MGS-TES surface temperatures for the ~1 hr difference in time-of-day between Mars Global Surveyor overflight and MRO overflight. These MGS-TES surface temperatures have 3°×7.5° spatial resolution in latitude and longitude (*M. Smith et al.*, 2004). The use of surface thermometry from MGS-TES is a proxy for more advanced surface thermometry, which will be fully incorporated later into the Lambert-albedo retrieval system for CRISM multispectral data. Enhancements to the spatial resolution of the surface temperature data will include: (i) an empirical approach to estimate the albedos between 3.8-3.92 μm based upon an observed correlation between albedos at 2.5 μm and albedos at 3.8-3.92 μm[12] (purple lines in Figure 1); and (ii) a physical approach to estimate the surface temperature for each pixel based upon using the ADR TE[13] lookup table which was computed from a physical thermal model of the martian surface (red and green lines in Figure 1). The empirical thermal-correction approach (*Jouglet et al.*, 2006) will allow the thermal emission to be estimated at 3.8-3.92 μm, which will facilitate the estimation of temperature for each CRISM spatial pixel, allowing thermal correction for shorter thermal wavelengths, i.e., between 3.0-3.8 μm. The physical thermal-correction approach (*Martin,* 2004*; Kieffer et al.,*1977) has input parameters that include surface albedo, solar incidence angle, surface slope, surface slope azimuth, altitude, and aerosol optical depths. Both of these more advanced thermal-correction techniques will allow better identification of $H_2O$ and $CO_2$ ices on the surface in minor quantities, for example.

---

[11] Ancillary Data Record -- Climatology
[12] Therefore, this empirical approach will need to wait until the CRISM data calibration is improved for wavelengths > 3.7 μm.
[13] Ancillary Data Record -- Thermal Emission

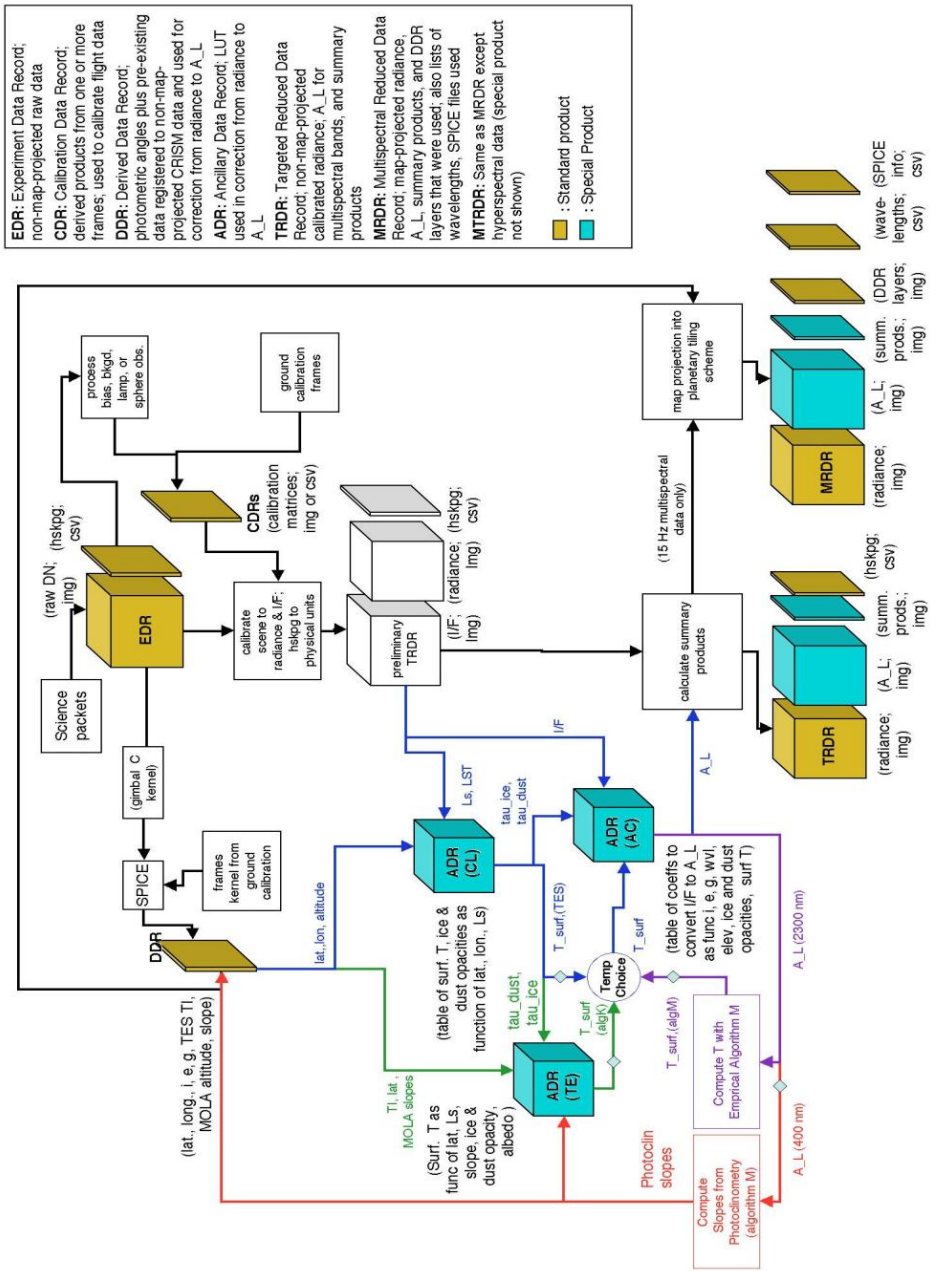

**Figure 1:** The software flowchart (*Murchie et al.*, 2007a) for converting the raw data (in the EDR) to calibrated *I/F* data (in the preliminary TRDR) to the photometrically, atmospherically, and thermally corrected Lambert albedo data (in the $A_L$ TRDR). The DDR is used for backplane information such as the photometric angles and the georeferencing latitudes and longitudes. The ADRs are used as lookup tables (LUTs) to query during the atmospheric and thermal correction; these ADR LUTs connect $A_L$ to radiance (and to I/F), i.e., the LUTs connect actual surface properties to spacecraft measurements. The MRDR[14] is computed by map-projecting a number of Lambert albedo TRDRs into a given map tile. In phase one of this project, we use climatological thermometry of the surface from the MGS-TES spectrometer. Therefore, the complexity of the thermal correction shown in this flow-chart is reduced considerably. In this paper, therefore, we focus particularly on the blue lines.

---

[14] Map-projected Reduced Data Record

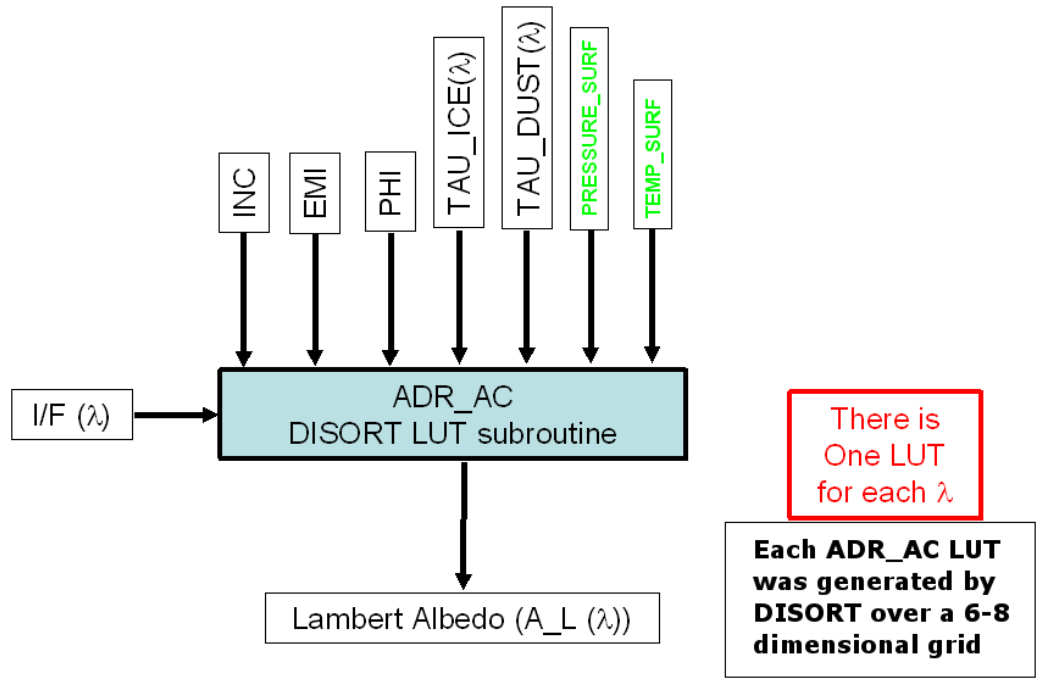

**Figure 2 :**
Detail of how the ADR AC is queried. This lookup table (LUT) is shown in context in figure 1. The ADR AC consists of $I/F(\lambda)$ values computed with forward calls for each grid point at several input values of the photometric angles (INC, EMI, PHI), several input values of the aerosol opacities ($\tau_{dust}(\lambda)$, $\tau_{ice}(\lambda)$), and several input values of Lambert albedo $A_L(\lambda)$. The surface pressure and surface temperature inputs are only included as input values to the grid forward-call computation for particular wavelength bands. For use in the CRISM multispectral retrieval of Lambert albedos, the ADR AC LUT is used in the inverse direction together with multidimensional interpolation, in order to use the $I/F(\lambda)$ value to estimate the Lambert albedo value, $A_L(\lambda)$.

## *3. DISORT-based radiative transfer retrieval of Lambert albedos*

In order to estimate the Lambert albedo for a givenCRISM spectel[15], we employ a lookup table (LUT) named the "Ancillary Data Record - Atmospheric Correction", or ADR-AC for short. This ADR-AC LUT was computed by using a DISORT-based[16] radiative transfer forward model software package (e.g., *Stamnes et al.*, 1988) to model how *I/F* will change as a function of 6-8 input parameters (see Figure 2). For a given spectel at given wavelength, λ, the subroutine for accessing the ADR AC LUT is called multiple times in order to 'invert' the LUT and to estimate the Lambert albedo from *I/F*. These multiple different values of Lambert albedo are then used for populating the values of the grid-point corners of a multi-dimensional cube. Multi-dimensional interpolation is then employed to estimate the value of Lambert albedo for interior points that are not grid points. In this way, Lambert albedos can be calculated for all spectels on in a multi-spectral map tile or image.

The use of these ADR AC LUTs allows a substantial acceleration in the retrieval of the Lambert albedos than would be allowed by direct forward calls to DISORT, reducing the computation time per MSP[17] TRDR strip (2700 rows, 60 columns, 72 spectral channels[18]) on a fast Linux or MSWindows workstation from about 3000 hours[19] to about 1 hour. This is much slower than the calibration software that converts from the raw data to *I/F*. It is also much slower than a simpler alternative Lambert-albedo retrieval software that corrects the *I/F* data for photometric effects by dividing by cos*(INC)* and for atmospheric effects using the 'Volcano-Scan' technique, without correcting for aerosol effects (for a discussion of some aspects of the Volcano-Scan technique, see subsection 5.b of this paper, as well as *Langevin et al.* (2005) and *Murchie et al.* (2007b)). Nonetheless, the benefits of quantitative accuracy for this technique, which is based upon radiative transfer and utilizes a physical model for the correction of effects due to aerosols and for absorption in the $CO_2$ gas bands, can be worth the extra computation time when compared to the Volcano-Scan technique.

After the application of the atmospheric correction software to each TRDR, the TRDRs are mosaicked into map tiles (Figure 1). At the equator, the tiles are of order 300km × 300km in size, corresponding to 5° × 5° in latitude and longitude. There will sometimes be more than 30 TRDR strips per map tile. With 1964 map tiles to span the planet, and with about 61% CRISM coverage of the planet[20] to date (as of March 1, 2008, with complete coverage expected), we need to make this Lambert albedo retrieval software as fast as possible, without sacrificing accuracy. Table Ia shows details on the grid parameters while Table Ib provides detail of the atmospheric layers currently being used for the CRISM multispectral retrieval of Lambert Albedos. The grid point spacing and atmospheric layer spacing can be adjusted in future deliveries in order to give more accuracy or higher speed, but these settings have thus far been adequate for both accuracy and speed.

---

[15] We use the definition of spectel as the contents of a spectral channel for a particular pixel. For example, I/F spectel #1 is the value of I/F for spectral channel #1; spectral channel #1 is the channel of the L detector of CRISM which is at λ = 3.92 μm.

[16] DIScrete-Ordinates-method Radiative Transfer

[17] Multispectral

[18] This image size roughly corresponds to 548km × 13km of surface area on Mars. The image width depends greatly on range to target. For the northern plains, the image width is about 13 km, but the MSP strips in the southern hemisphere are much closer to 10 km wide.

[19] Using 1 second for computing the Lambert albedo via forward calls for each spectel, this gives
        (2700*60*72 spectels)*(1 sec/spectel)/(3600 sec/hr) = 3240 hrs.

In Figure 3, we show a portion of the internal structure of the ADR AC LUT#2007. This LUT was computed for the deep atmospheric $CO_2$ gas band at 2.007 μm. Note the straight line dependence on semi-log plots. This indicates that if we use exponential interpolation in the surface-pressure variable, that the non-linear dependence of *I/F* upon surface pressure can be adequately sampled by only 3 grid points for the pressure data axis. The exponential pressure scale-heights of *I/F*, which we denote as $p_{sh;I/F}$, range from ~5 mbars to ~12 mbars for different values of the dust and ice optical depths. Though we don't show the dependencies of *I/F* on the other 5-7 data axes here, the majority of the parameter spaces for the other axes are dominated by nearly linear dependencies. Because these dependencies are not strictly linear, and occasionally non-monotonic, we sample most of the other 5-7 data axes with more than 3 grid points. The plots in Figure 3 also demonstrate that differences in ice and dust aerosol optical depths do have quite an effect on the coefficients to the exponential dependence of *I/F* on surface pressure. The use of multi-dimensional ADR AC LUTs[21] allows much faster computation than forward calls, and more accuracy than limited analytical or empirical models. This combination of speed and accuracy provides sufficient flexibility and robustness for the creation of the set of global map tiles of Lambert albedo for Mars.

---

[20] The coverage depends on latitude: for the equatorial region, the coverage is 47%; the poles are nearly complete. Furthermore, this coverage does not account for periods of low data quality, most notably during periods of high dust optical depths in the atmosphere.

[21] The ADR AC LUTs have sizes: 13.1 MB for each of the thermal bands; 4.4 MB for each of the $CO_2$ gas bands; and 1.5 MB for the each of the normal bands. The sizes of these LUTs have been minimized to the extent possible, in order to optimize both the speed and the usage of memory.

| Variable | # of grid points for data axis | Min. value | Max. value | Spacing |
|---|---|---|---|---|
| *The six grid axes that are common to all CRISM multispectral bands* | | | | |
| Cos(EMI) | 7 | 0.10 | 1.00 | 0.15 |
| PHI (deg) (azimuthal angle) | 11 | 0.00 | 180.00 | 18.00 |
| Cos(INC) | 7 | 0.10 | 1.00 | 0.15 |
| $A_L$ | 11 | 0.00 | 0.60 | 0.06 |
| $\tau_{dust}$(9.3 μm) | 9 | 0.01 | 0.71 | 0.0875 |
| $\tau_{ice}$(12.1 μm) | 7 | 0.00 | 0.50 | 0.0833 |
| *Extra DISORT grid axis used only for $CO_2$ gas bands:* *These gas bands are at: 1.21, 1.43, 1.66, 1.88, 1.97-2.17, 2.60, 2.70, and 3.00 μm* | | | | |
| Surf Press.(mbar) | 3 | 1.00 | 8.00 | 3.50 |
| *Extra DISORT grid axis used only for thermal bands (3.12-3.92 μm):* | | | | |
| Surf. Temp.(K) | 3 | 180.00 | 300.00 | 60 |
| *Extra DISORT grid axis used only for thermal bands (3.12-3.92 μm):* *The limits for two example thermal bands are shown (3.12 μm and 3.92 μm).* | | | | |
| Solar Irradiance in W/m²/μm @3.12 μm | 3 | 0.1113 | 0.1855 | 0.0371 |
| Solar Irrad. in W/m²/μm @3.92 μm | 3 | 0.0526 | 0.0877 | 0.0175 |

### Table Ia:

The definition of the grid points used by DISORT to generate the ADR AC LUTs. For each of the grid points, DISORT is run in order to estimate *I/F*. The CRISM_LambertAlb system later uses the ADR AC LUT for each wavelength band together with multi-dimensional interpolation to estimate the Lambert Albedo ($A_L$) from the measured *I/F* value and the estimated values of the other data axes (the photometric angles: EMI, PHI, INC; the aerosol opacities: $\tau_{dust}$, $\tau_{ice}$; and the Surface Pressure, the Surface Temperature, and the Solar Irradiance). Three of the data axes for the grid definition are only used when either the spectral band is a $CO_2$ gas band or a thermal band.

| *Number of DISORT moments* | *Number of DISORT streams* | *Number of atmospheric layers* |
|---|---|---|
| 64 | 32 | 10 |

| *Heights of boundaries between layers (km)* | | | | | | | | | | |
|---|---|---|---|---|---|---|---|---|---|---|
| 50.0 | 40.0 | 30.0 | 25.0 | 20.0 | 15.0 | 10.0 | 5.0 | 2.5 | 1.0 | 0.0 |

**Table Ib:**

Vital numbers for the atmospheric model used by DISORT in order to compute the ADR AC LUTs. For the atmospheric equation of state, we use a representative temperature profile from the MGS-TES database and simply integrate the hydrostatic structure equation. Given the negligible contribution from atmospheric thermal emission and the minimal sensitivity of the molecular absorption to the expected excursions in atmospheric temperatures, our use of a representative profile is quite adequate. We also used the Lambertian phase function (i.e., instead of a Hapke phase function) for the surface photometry. For the $CO_2$ gas bands, we also employ the correlated-k technique (for a general description, see *Thomas & Stamnes* (1999)).

As we will discuss in the next section on climatology, we use the estimates for the dust and ice aerosol optical depths from the TES instrument on MGS as inputs to the DISORT-based ADR AC LUT, in order to estimate the Lambert albedo spectra from the *I/F* spectra. However, the MGS-TES instrument measures the optical depths of the dust and ice aerosols at mid-infrared wavelengths: 9.3 μm for dust aerosols, and 12.1 μm for ice aerosols. We have employed a Mie-scattering approximation (assuming spherical aerosols) (*Wolff & Clancy*, 2003) in order to convert the extinction cross-sections of the aerosols at MGS-TES mid-infrared wavelengths to the extinction cross sections at CRISM wavelengths (0.362-3.92 μm)[22]. This normalized extinction cross-section is used multiplicatively to convert the aerosol optical depths measured at TES wavelengths to the aerosol optical depths measured at CRISM wavelengths. In order to generate scattering

---

[22] The extinction cross-section is defined as being the sum of the absorption cross-section and the scattering cross-section. The cross-section of a particle is roughly equivalent to the effective area (i.e., of scattering or absorption) that a particle or a molecule presents to an incoming photon.

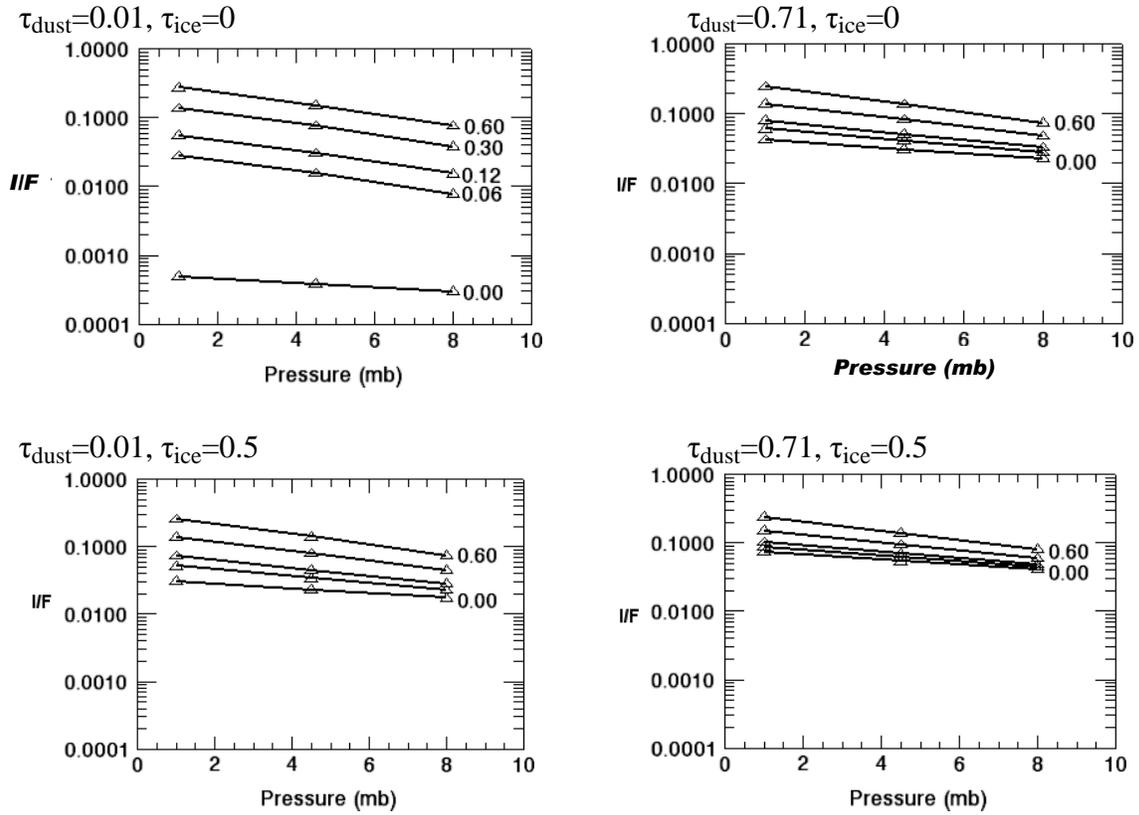

## **Figure 3:**

We plot the dependencies of *I/F* on surface pressure in semi-log plots, in order to show a portion of the internal structure of the ADR AC LUT, for $\lambda = 2.007$ μm. These dependencies are given for 5 different values of Lambert albedo ($A_L$ = 0.0, 0.06, 0.12, 0.30, 0.60) and four different sets of dust and ice aerosol optical-depth values. The indicated aerosol optical depths are for the MGS-TES reference wavelength. The straight line semi-log dependencies suggest an exponential dependence, as is confirmed by detailed forward calls to DISORT (not shown here). The wavelength $\lambda=2.007$ μm is deep in the $CO_2$ gas band. These dust and ice optical depths are the values at 2.007 μm, not the values for dust and ice optical depth at the mid-infrared reference wavelengths. For the input data axis of surface pressure, we find that 3 grid points form adequate sampling intervals from 1-8 mbars of surface pressure, but only if we use exponential interpolation for this data axis.

properties, for ice aerosols, we use the formulation from *Warren* (1984) with $R_{eff} = 2.0$ μm and variance = 0.1 μm, and for dust aerosols, we use $R_{eff} = 1.7$ μm and variance = 0.4 μm. For further discussion, see *Clancy et al.* (2003).

## *4. Pressure, Temperature and Aerosol Climatology*

In order to serve as our initial inputs to the DISORT-based Lambert-albedo retrieval system, we are using the climatological records of aerosol optical depths and temperature from MGS-TES (*M. Smith*, 2004), as well as the surface altitudes from laser altimetry by the MOLA instrument (*D. Smith et al.*, 2001), which is also on MGS. The MGS-TES climatology is in a LUT called "Ancillary Data Record – Climatology" (ADR CL), and for our initial work we are using the MGS-TES climatology from Mars years 24 and 26 (1999-2000, 2003-2005).[23] The MOLA altimetry is given for each pixel in the TRDR in a separate Derived Data Record (DDR).

The MGS-TES surface temperature is used as an input to query the DISORT-based ADR AC LUT in order to retrieve Lambert albedos for the thermal bands at wavelengths >3.0 μm. The optical depths for dust and ice aerosols are used by DISORT-based ADR AC LUT to retrieve Lambert albedos for all spectral bands, but most of the effect is at wavelengths <1.2 μm. The MGS-TES temperature of the lower atmosphere is used together with the MOLA altimetry and a Viking-based climatological record of surface pressure (*Tillman et al.*, 1993), in order to estimate surface pressure for each spatial pixel in the TRDR, for different locations on the planet at different times in the Mars year. The MGS-TES lower-atmospheric temperature is computed by averaging over the four lowest-available MGS-TES levels (above the surface), which is equivalent to averaging over 1 atmospheric pressure scale height.

In Figure 4, we show the MGS-TES climatological record of the surface pressure, the temperatures, the aerosols and the water vapor for two longitudes and three solar longitudes[24] as function of latitude. This figure demonstrates the climatological spatiotemporal variability that is encapsulated in the ADR CL LUT. Such variability implies two things: (i) using a simple analytical or empirical model of most of these climatological variables is not sufficient for our purposes of retrieving Lambert albedos (hence, the importance of the ADR CL LUT); and (ii) with such past variability, it is unlikely that the ADR CL LUT will always be correct (hence, the need for direct measurements of the climatological variables from the CRISM data). A secondary intent of this figure is to demonstrate that there is missing climatological data in the ADR CL LUT. The absence of data is indicated in part by the dashed horizontal lines in these traces, particularly near the poles. The data value for the missing data locations and times is chosen to be the value for the nearest available location and time. Often, these nearest replacement values are inadequate for the accurate retrieval of Lambert albedos. This is particularly relevant for the missing aerosol values (i.e., near the poles). The aerosol values near the poles can be quite high, and somewhat variable (*Titus and Kieffer*, 2001; *M. Smith et al.*, 2001; *M. Smith,* 2004; *Newman, Lewis, Read*, 2005; *Zasova et al.*, 2005), and if their values are inferred from locations further from the poles, highly inaccurate retrievals of Lambertian albedos will ensue. Hence, we map the regions of missing aerosol data or poor aerosol data quality in Figure 5. This map shows for example that the CRISM north polar mapping campaign will likely require, for accuracy, the direct retrieval of aerosol abundances instead of the nearest MGS-TES aerosol optical depths.

---

[23] The use of Mars years 24 and 26 is part of the ADR_CL operating mode named 'Year of Best Data Quality'. Mars year 24 has the best data quality for TES, but Mars year 24 starts at $L_s = 103°$. The data prior to $L_s = 103°$ is coming from Mars year 26, since Mars year 25 had a dust storm.
[24] Solar longitude ($L_s$) ranges from 0-360° and represents the position of Mars with respect to the Sun. The solar longitude of $L_s = 0°$ corresponds to the spring equinox in the northern hemisphere on Mars.

In the wavelength range 1.2-2.6 μm, the surface pressure is the most crucial parameter for correcting the data. And there are subtleties in estimating the surface pressure climatologically (*Tillman et al.,* 1993; *M. Smith*, 2004; *Spiga et al.*, 2007; *Forget et al.*, 2007), hence the emphasis in the discussion in this section on the surface pressure. The aerosols are primarily active shortwards of 1.2 μm, but they do have low spectral-frequency effects at longer wavelengths, as well as interaction effects with the $CO_2$ gas bands at 2 μm. The aerosols are determined directly from the MGS-TES climatological LUT (ADR CL). Therefore, with the exception of the rescaling as described in *Wolff & Clancy* (2003), and the data-quality constraints shown later in Figure 5, we limit the amount of discussion here for the aerosols, relative to the amount of discussion for surface pressure. Surface temperature for thermal correction longwards of 3 μm is also given directly by MGS-TES climatology in the ADR CL LUT. Surface temperature has similar data quality constraints as the aerosols do. Therefore, we defer more detailed discussion of the aerosols and surface temperatures inputs to the DISORT correction until a later phase in the CRISM multispectral Lambert albedo retrieval project when we start retrieving aerosol optical depths and surface temperatures directly from the CRISM data.

## 4.1 Algorithmic Pressure Climatology

The surface pressure on Mars varies seasonally (cf. *Tillman et al.,* 1993) and we must encapsulate that variation in an analytical, 'algorithmic' expression. This algorithmic climatology for surface pressure used by CRISM_LambertAlb is based upon an average of measurements of the two Viking landers over two Mars years without great dust storms (Viking Lander 1 for two Mars years and Viking Lander 2 for one Mars year, see *Tillman et al.,* (1993)). The technique adopted by the CRISM team for averaging of the data from the two landers is detailed below, and is similar to unpublished techniques used by the MGS-TES team. The pressure cycle over the course of a Mars year has characteristically two large and broad peaks, roughly preceding the winter solstice for each of the polar caps, after which $CO_2$ condensation on the caps begins in earnest. The Mars-year averaging is done for each of the 5 harmonic amplitudes and phases, and the averages are shown in Tables IIa, IIb, & IIc. This pressure climatology as a function of solar longitude or Julian day is then offset to correspond to a MOLA altitude of 0 km. The 'algorithmic' Viking-based climatological pressure equation used for CRISM multispectral retrieval of Lambert albedos is:

$$p_{\text{surf}}(f, z=0 \text{ km}) = p_{\text{surf,ave}}(z=0 \text{ km}) \times (1 + (1/p(\text{VL})) \Sigma_k [a_k \sin(2\pi k f + \varphi_k)] \quad (1)$$

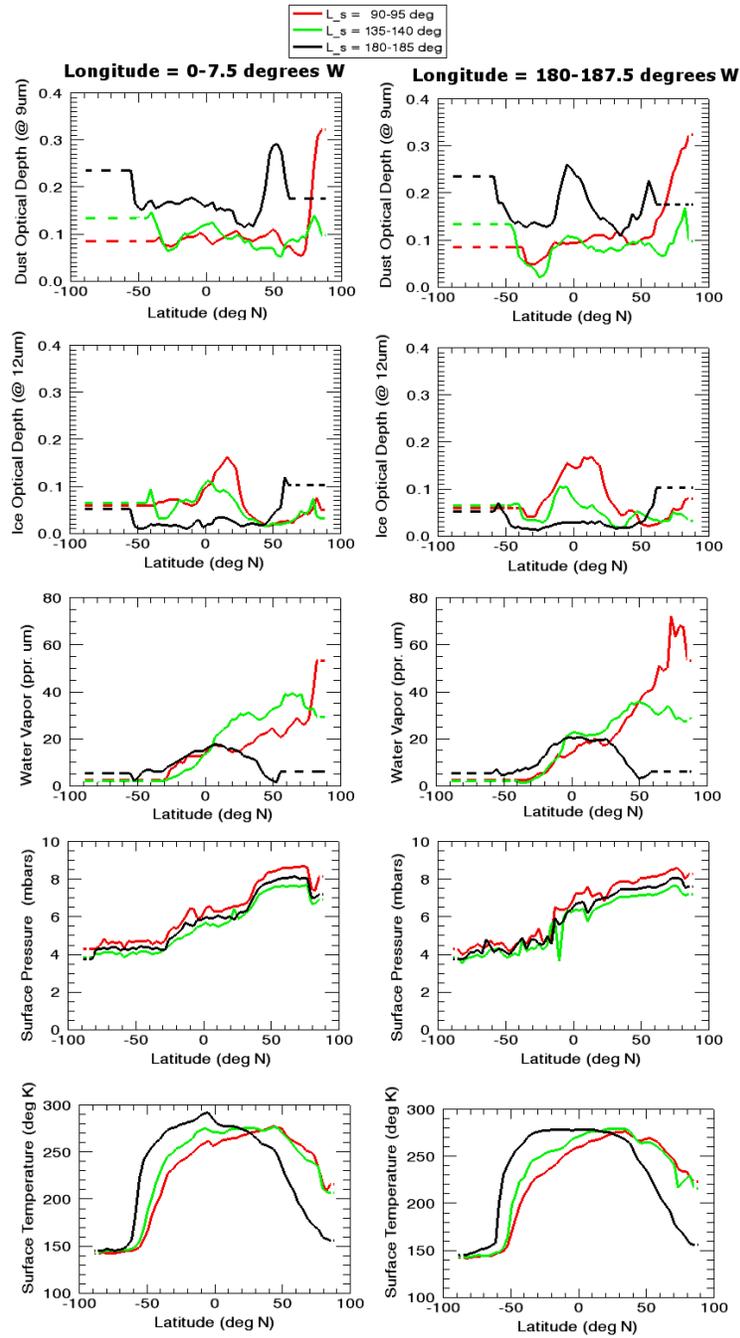

## **Figure 4:**

ADR CL climatological profiles as a function of latitude, for the indicated longitude ranges and the indicated solar longitude ranges. The dust and ice aerosol optical depths are shown at the mid-infrared MGS-TES reference wavelengths. Neither the ADR CL low spatial-resolution surface pressure nor the MGS-TES-estimated water vapor, both shown here, is used in the CRISM_LambertAlb system. We use the MGS-TES-estimated temperature of the lower atmosphere (not shown here), together with MOLA altimetry and Viking pressure climatology in order to 'algorithmically' estimate surface pressure at nearly the CRISM pixel scale. The MGS-TES-estimated surface temperature is used for the thermal correction for wavelengths >3.0 μm. The dashed horizontal line segments represent gaps in the data (due to missing data or sub-optimal data quality) that are currently filled in with nearby data.

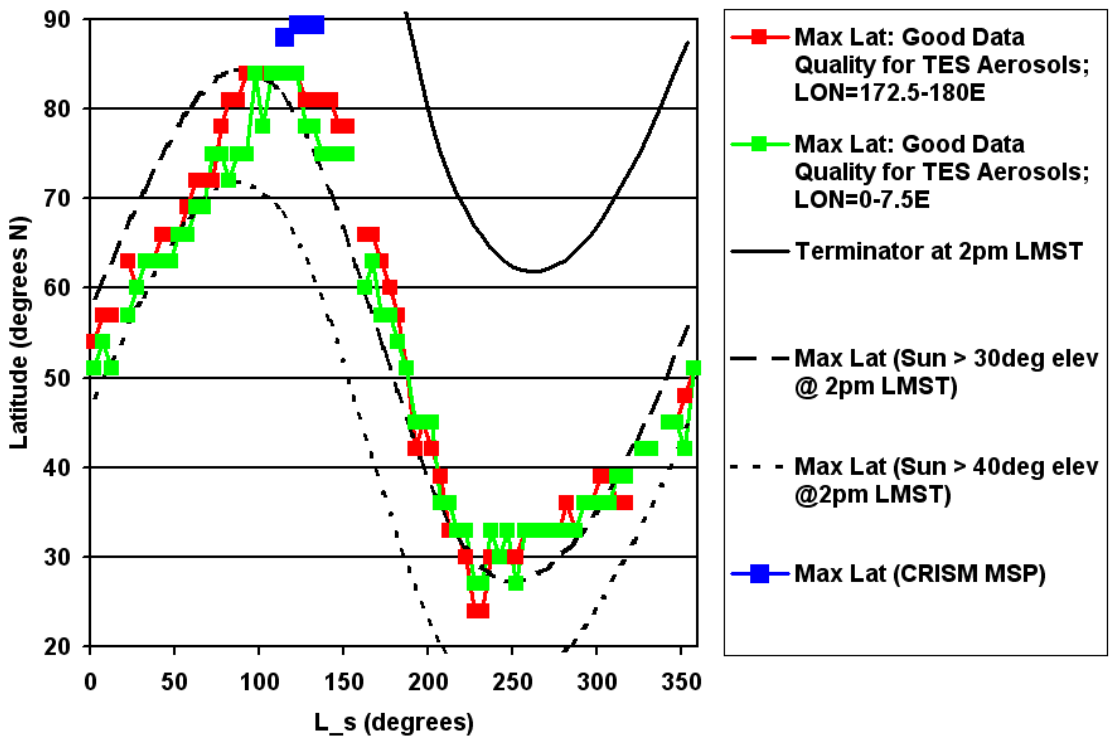

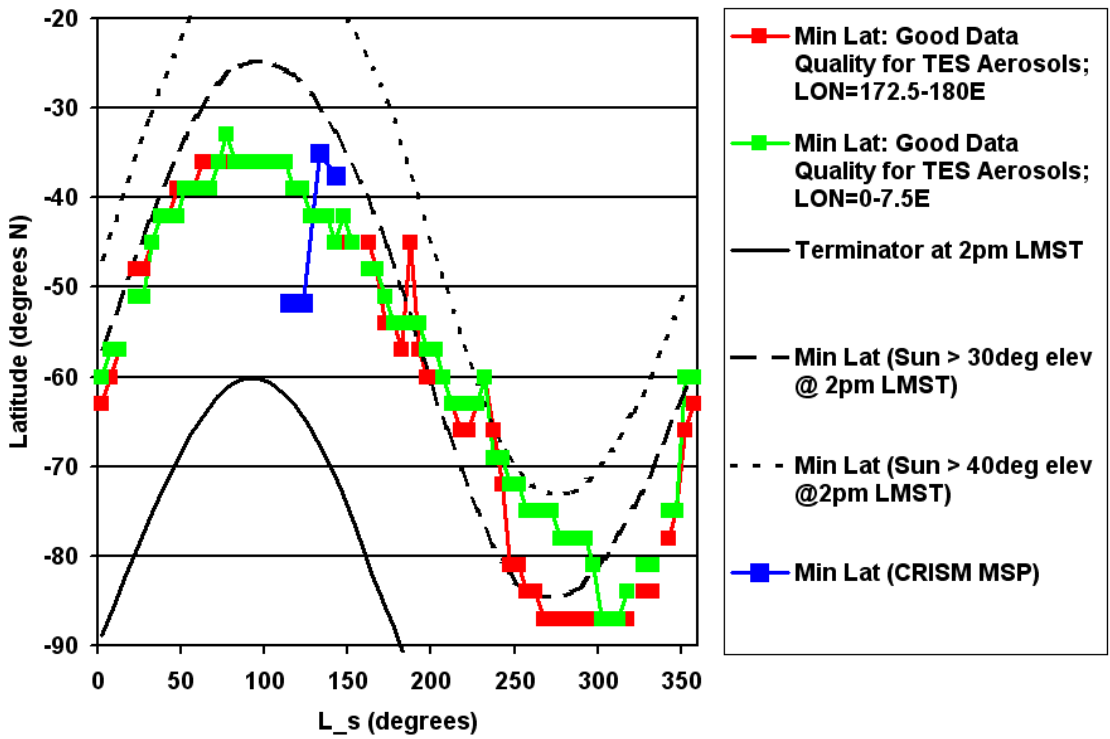

**Figure 5 (previous page):**
We plot the latitudinal constraints of good MGS-TES-based aerosol climatology as a function of solar longitude ($L_s$) for two different planetocentric longitude ranges (0-7.5E and 172.5-180E), in order to show the northern-most and southern-most bounds for current phase I atmospheric correction of CRISM multispectral data. Temporal gaps in the aerosol climatology are shown by gaps in the green and the red curves. Also plotted are the positions of the terminator and the constraints on solar incidence angle as a function of solar longitude ($L_s$). The maximal extents of acquisition of CRISM multispectral data are also indicated, for the data that is currently available (August 2007) in the PDS geosciences data archive. Note that the maximal extents of MGS-TES aerosol data quality correspond roughly to the range of >30-40° incidence angle for much of the martian year.

In this equation, $p_{surf,ave}(z=0$ km$)$ corresponds to the average surface pressure over a martian year for a surface that is at a MOLA altitude of z=0 km, and $p_{surf}(f, z=0$ km$)$ is the time-dependence of this surface pressure. The values of the amplitudes, $a_k$, and the phases, $\varphi_k$, are the averages in tables IIb,IIc, the phases being given in radians. Seasonal time is expressed as $f$ in terms of fraction of a martian year *in seconds, not in $L_s$* (i.e., $f$ varies from 0 to 1). Note that $L_s$ is *not* linear in time, so we use calendar date (or spacecraft clock time). In the above equation, $f$ is equal to zero at $L_s$=330.2°, as in *Tillman et al.* (1993). Relevant calendar dates for $L_s$=330.2° include: Nov. 26, 2005 (Julian Date 2453701) and Oct. 14, 2007 (Julian date 2454388).

The average for the 3 mean pressures is 8.180 mbar, but this is for the mean elevation of the Viking landers. The value of $p_{surf,ave}(z=0$ km$)$=5.477 mbar that is in our formulation is determined by adjusting the $p$(VL)= 8.180 mbar mean value for the elevation of the Viking landers.

The pressure at other altitudes is computed from an exponential dependence with a given scale height, $H$:
$$p_{surf}(f, z) = p_{surf}(f, z=0 \text{ km}) \exp(-z/H), \quad (2)$$
where the scale height depends on the temperature of the lower atmosphere: $H = RT/g$. The gas constant is $R$; the gravitational acceleration is $g$; and the lower atmospheric temperature is $T$ (in Kelvin). For Mars, $H = (T / 19.5)$ km, which for most daytime non-polar conditions is roughly 10 km. The temperature of the lower atmosphere for a given location and time is determined from the MGS-TES climatology described earlier.

Using Eq. 1, we have calibrated for atmospheric conditions at the locations of the Viking landers 25-30 Earth years ago. This should allow somewhat accurate pressure determination at most locations on the planet, if we only know the altitude and the temperature of the lower atmosphere. This goes beyond what can be accomplished with the 3° × 7.5° (in latitude and longitude) spatial resolution and 5° (in $L_s$) temporal resolution that we have available in the ADR CL LUT. The surface pressures in the ADR CL LUT are also based upon Viking climatology, but the ADR CL uses a different definition of scale height than the one used in the algorithmic climatology here, and is sampled at much lower spatial resolution than the MOLA grid. The pressure climatology in the ADR CL is a rather crude estimate since it uses a constant 10 km for the scale height, and the climatology values are simply averages over the entire climatology bin from whatever observations happen to fall in that bin (which can be from very different altitudes than the multispectral pixel of interest). The differences between the two climatologies are usually less than 20%, and often better than this. The largest systematic differences between the ADR CL LUT climatology and the algorithmic climatology are observed in the Hellas basin, where the differences are greater than 1.5 mbars. A comparison is currently being performed of the algorithmic pressure climatology, based in a large part on data from the Viking Landers, and pressures retrieved directly from CRISM data, and which will be a topic in a future paper.

By inspection of Figure 3 with the exponential dependence of *I/F* on surface pressure of a certain *I/F* scale height, $p_{sh;I/F}$, and by some calculus, a +1.5 mbar error in pressure, for a $p_{sh;I/F} = 7.5$ mbar, will have a -20% effect on the measured *I/F*. Therefore, if we want 5% photometry in the gas bands for this particular case of $p_{sh;I/F} = 7.5$ mbar, we need 0.38 mbar accuracy in the pressure climatology. For certain locations on the planet, we will not be able to achieve this accuracy, one example being: 'in very rough terrain', where there is significant variation in elevation at spatial scales smaller than the MOLA footprint of ~300 m. However, the current accuracy in high spatial-resolution Viking-based algorithmic pressure climatology is adequate for many purposes and over much of the planet, and will serve CRISM data-processing needs for the time being, until we upgrade the Lambert-albedo retrieval system to retrieve the surface pressures directly from CRISM data.

Table IIa: Mean Pressure

| Mars year, Lander | Mean Pressure (mbar) |
| --- | --- |
| Year 1, VL1 | 7.936 |
| Year 2, VL1 | 7.942 |
| Year 1, VL2 | 8.663 |
| Ave = $p$(VL) | 8.180 |

Table IIb: Amplitude of first 5 harmonic terms of Pressure Cycle, averaged over the first two Mars years for Viking Lander 1, and the first Mars year for Viking Lander 2.

| $a_1$(mbar) | $a_2$(mbar) | $a_3$(mbar) | $a_4$(mbar) | $a_5$(mbar) |
| --- | --- | --- | --- | --- |
| 0.704 | 0.582 | 0.108 | 0.062 | 0.015 |

Table IIc: Phase of first 5 harmonic terms of Pressure Cycle, averaged over the first two Mars years for Viking Lander 1, and the first Mars year for Viking Lander 2.

|  | $\varphi_1$ | $\varphi_2$ | $\varphi_3$ | $\varphi_4$ | $\varphi_5$ |
| --- | --- | --- | --- | --- | --- |
| In Degrees | 92.31° | -130.80° | -69.76° | -10.00° | 49.55° |
| In radians | 1.611 rad | -2.283 rad | -1.217 rad | -0.175 rad | 0.865 rad |

## *5. Applications of the CRISM_LambertAlb system*

We show here a small subset of the results and analyses that we have completed thus far. The intent of these analyses is to elucidate some of the capabilities of the CRISM_LambertAlb system first described in *McGuire et al.* (2006) for the correction of multispectral CRISM data. The results that we describe here are: (a) comparison of Lambert albedo spectra from DISORT retrieval with library reflectance spectra of minerals (b) analysis of corrected spectra from near the northern ice cap and comparison of these corrected spectra with the 'Volcano-Scan' technique; (c) intercomparison of map tiles in the Tyrrhena Terra region northeast of the Hellas basin; and (d) demonstration of map-tile production for a region north of the Phoenix Lander 2007 landing site.

The CRISM instrument contains two detectors, one for wavelengths from 0.362-1.04 μm (named the S detector), and another for wavelengths from 1.02-3.92 μm (named the L detector). The *I/F* datacubes for all of the analyses described here use calibration 'TRR2', which is the current calibration as of September 2007. The version of the volcano-scan software is CAT 6.0, which uses cos(INC) photometric correction, together with a transmission spectrum of the martian atmosphere derived from taking a ratio of a TRR1 version of the *I/F* spectrum at the summit to the I/F spectrum at the base of Olympus Mons. The version of the CRISM_LambertAlb system is 20070809, which is the latest version.

CRISM_LambertAlb does not correct for atmospheric water vapor or carbon monoxide. These corrections can be done afterwards with special runs of DISORT (which we do not present here), and they are perhaps best done after the main pipeline processing because:
1. Water vapor and carbon monoxide are relatively minor species in the atmosphere. Twenty precipitable μm of water vapor is a typical column-depth of water, and 700 ppm of carbon monoxide is typical for the martian atmosphere (*M. Smith*, 2007).
2. The band-depths and band-widths of water vapor and carbon monoxide are relatively small compared to those of carbon dioxide.
3. Water vapor is somewhat variable in abundance even on small spatial scales, so the amount of water vapor may need to be inferred directly from the CRISM data on a pixel-by-pixel basis.

### 5.a) Comparison of Lambert albedo spectra from DISORT retrieval with library spectra of minerals

The CRISM mapping strip MSP00002838_07, in the far northern latitudes near the northern ice cap has some nice spectral and mineralogical diversity, which allows the quick analysis of the correction capabilities of CRISM_LambertAlb. The mineralogical diversity includes: $H_2O$ ice, gypsum-bearing dunes and dust-covered surfaces. This mineralogical diversity gives rise to spectral diversity that includes bright spectra, dark spectra and spectra that are bright at some wavelengths and dark at others. In Figure 6A, three different spectra from a single multispectral strip (MSP00002838_07) in the northern latitudes are shown. This strip was chosen because it has both icy outcrops and gypsum-rich dunes (*Roach et al.*, 2007) within it.

In Figures 6B and 6C, we compare two Lambert albedo spectra from this strip with the USGS reflectance mineral spectra (*Clark et al.*, 1993) for gypsum and water ice. We have also multiplied the CRISM spectra in these two figures by an arbitrary constant, in order to better compare with the library spectra; this can account for the effect of possible spectrally-bland darkening agents on the surface (*Clark*, 1983).

The CRISM MSP Lambert albedo spectrum in Figure 6B has a strong absorption at 1.9 μm, and an absorption edge at 2.4 μm, which is a hallmark of polyhydrated sulfate minerals (*Murchie et al.*, 2007b). The twice-hydrated mineral gypsum compares somewhat well with this CRISM spectrum, including the absorptions at 1.9 and 2.25 μm, and the drop-off at 2.4 μm (*Roush et al.*, 2007). However, the CRISM data lacks an absorption at 1.8 μm, and the 1.45 μm band for gypsum correspond to a band at slightly longer wavelengths in the CRISM data. This CRISM spectrum may not correspond to gypsum, but with good certainty, it corresponds to a sulfate-rich mineral.

The case is much stronger in Figure 6C that the CRISM spectrum corresponds to an area that is rich in $H_2O$ ice, with the absorptions at 1.3, 1.5, 2.0, and 2.5 μm (*Clark and Roush*, 1984; *Warren*, 1984; *Grundy and Schmitt*, 1998; *Painter et al.*, 1998; *Green et al.*, 2006).

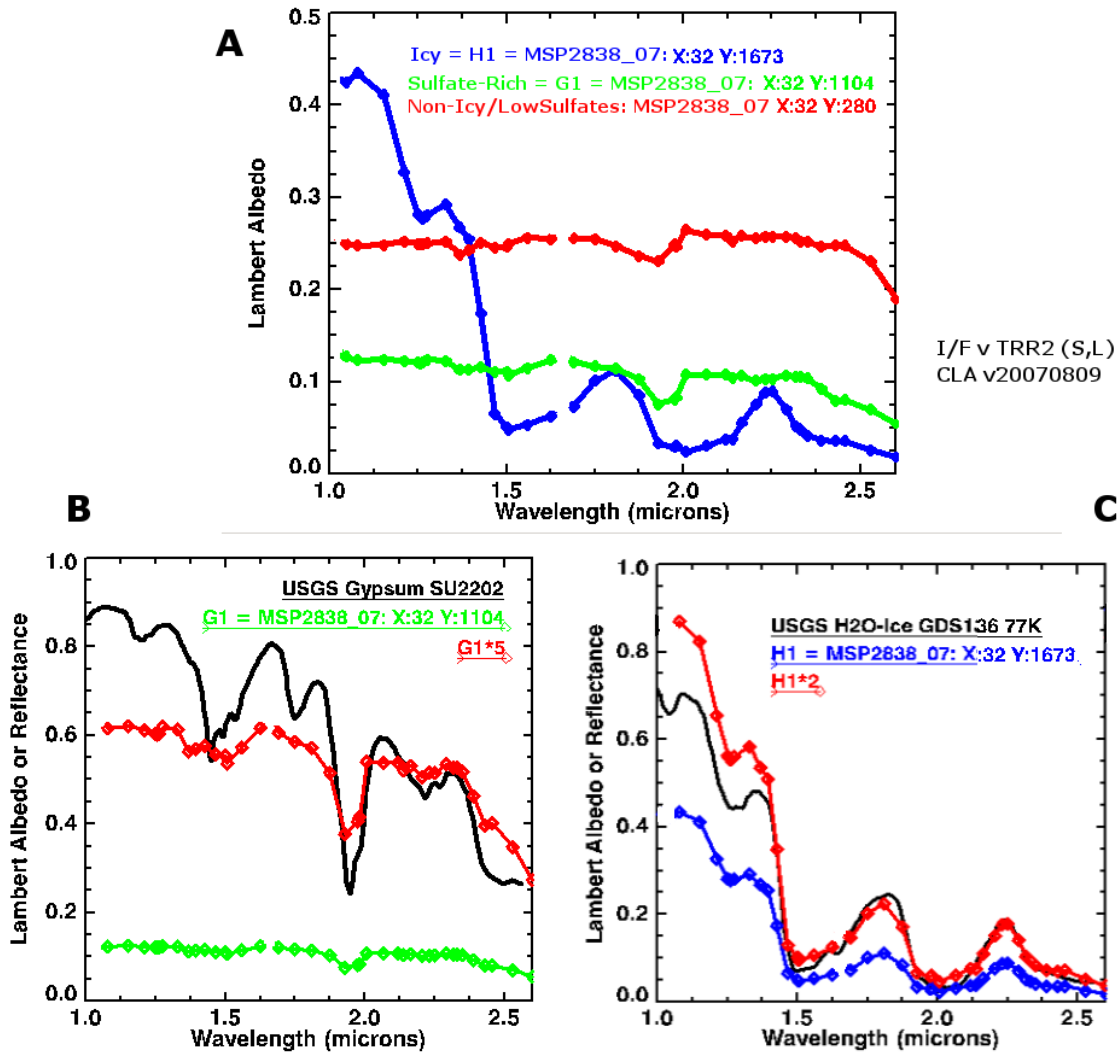

## **Figure 6:**

In (A), three Lambert albedo spectra from CRISM multispectral strip MSP00002838_07. This strip is located near the north pole and crosses the gypsum dunes in Olympia Undae. These Lambert albedo spectra were computed using the CRISM_LambertAlb software system. In (B), we show the sulfate-rich spectrum from (A), together with a Gypsum spectrum from the USGS library of reflectance spectra of minerals (*Clark et al.*, 1993). In (C), we show the ice-rich spectrum from (A), together with a $H_2O$-ice spectrum from the USGS library of reflectance spectra of minerals. Both (B) and (C) also contain versions of the CRISM spectra that are scaled by multiplying by an arbitrary constant, so that the CRISM spectra can be better compared with the USGS library reflectance spectra. The X and Y values in each sub-figure are the locations of the pixel in CRISM multispectral image MSP00002838_07. For clarity, known channels of poor data quality (wavelengths of 1.02, 1.05 and 1.65 μm) have been removed from these spectra.

## 5.b) Comparison of DISORT correction and Volcano-Scan correction

Such spectral diversity allows a decent measure of testing of the overall quality of the atmospheric correction capabilities of our system, particularly in the presence of obscuring aerosols in the martian atmosphere. In Figures 7 and 8, we compare two different atmospheric-correction techniques, the DISORT-based climatological system described in this paper and the volcano-scan correction system (*Langevin et al.*, 2005). The volcano-scan correction system relies on a measurement of the atmospheric transmission by the CRISM instrument, accomplished by taking the ratio of a nadir-looking *I/F* spectrum acquired at the summit of Olympus Mons to the *I/F* spectrum acquired at the base of Olympus Mons. The volcano-scan correction technique rescales this transmission spectrum for variable surface pressure at other locations of the planet. The volcano scan technique does not account for aerosol optical depths that vary with time or location on the planet, nor does it correct *I/F* spectra well when there are ices present on the surface. Figure 7 shows this comparison between different atmospheric-correction techniques for bright spectra, and Figure 8 shows this comparison for icy spectra that are both bright and dark.

Note that the atmospheric correction works well for both techniques in the $CO_2$ gas bands at 2.0 μm for the bright spectra. However, for the icy spectra which are dark in the broad $H_2O$ ice bands at 1.5, 2.0, and greater than 2.4 μm, the DISORT-based method handles the $CO_2$ gas-band correction at 2.0 μm much more effectively. This latter deficiency of the Volcano-Scan technique (as currently implemented in CAT 6.0) is due to a combination of the unmodeled interaction of the aerosols at these dark wavelengths and the technique of fitting the band-depth at $CO_2$ when ices are present on the surface ($H_2O$ and $CO_2$ ices also have deep bands at 2.0 μm). Also, note that the more subtle differences between the two atmospheric-correction techniques over the entire spectral range from 1.0-2.5 μm in both figures. Again, this is due to the variable aerosols not being accounted for in the Volcano-Scan technique.

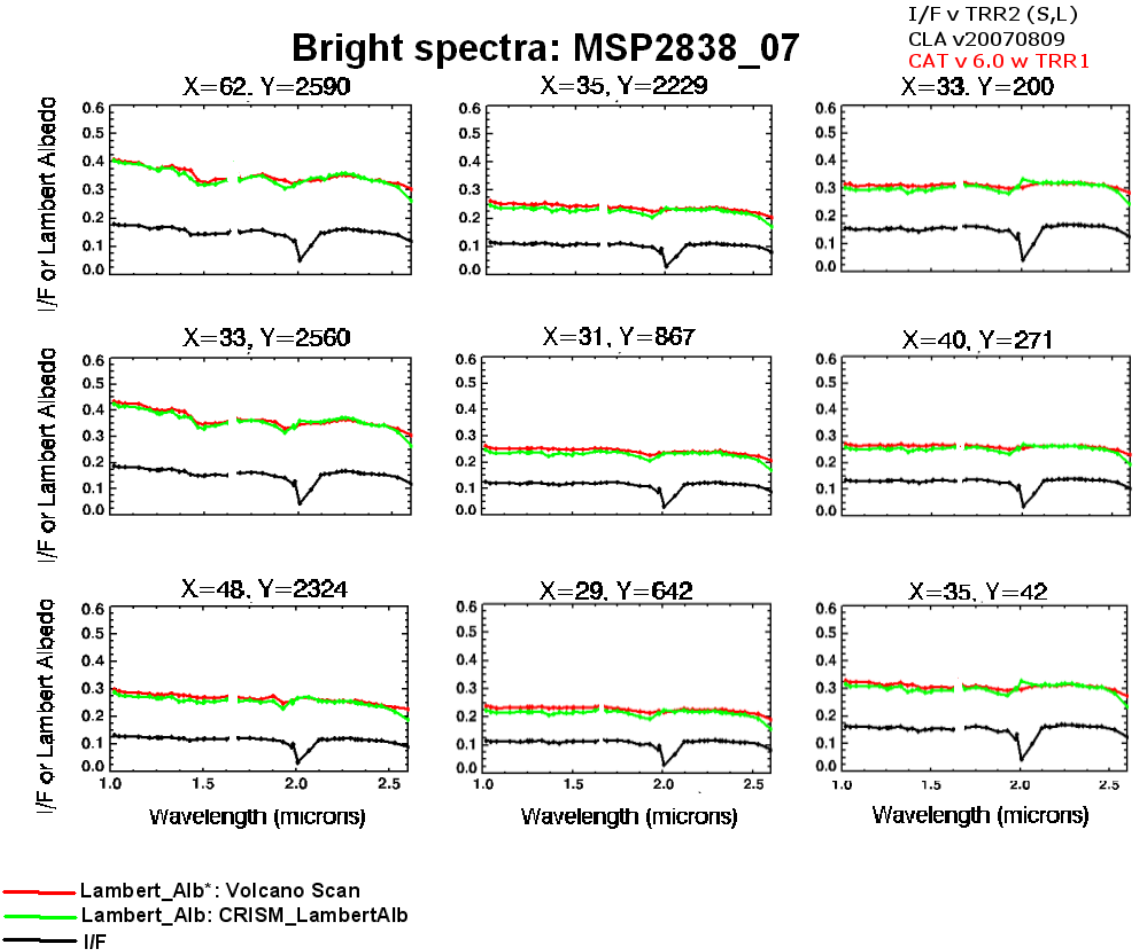

## Figure 7:

We show some of the brighter L-detector spectra from the CRISM Multispectral observation, MSP00002838_07, located in the northern polar region (approximately 78-85N, 240E), acquired on September 30, 2007 ($L_s$ =114.1 degrees). Only the wavelengths < 2.6 μm are shown here. The black traces are the measured *I/F* spectra; the red traces (Lambert_Alb*) are the quasi-"Lambert albedo" spectra atmospherically-corrected by the Volcano-Scan technique and photometrically-corrected by dividing by *cos(INC)*; the green traces (Lambert_Alb) are the Lambert albedo spectra atmospherically- and photometrically-corrected by the DISORT-based technique discussed in this paper. The X,Y coordinates in the multispectral image are indicated; X values near 32 are in the central portion of the 64-column image; Y values near 2700 are in the northern areas of the 2700-row image, where there is a moderate abundance of ice grains detectable in the spectra. For clarity, the artifact near 1.65 μm is due to a filter boundary in the CRISM instrument has been removed from these spectra. The deep atmospheric $CO_2$ absorption is well-corrected by both techniques. The offset in Lambert albedo between the two correction techniques (i.e., between 1-2 μm) is caused by aerosol optical depths being explicitly handled by the DISORT-based technique, and not handled by the Volcano-Scan technique.

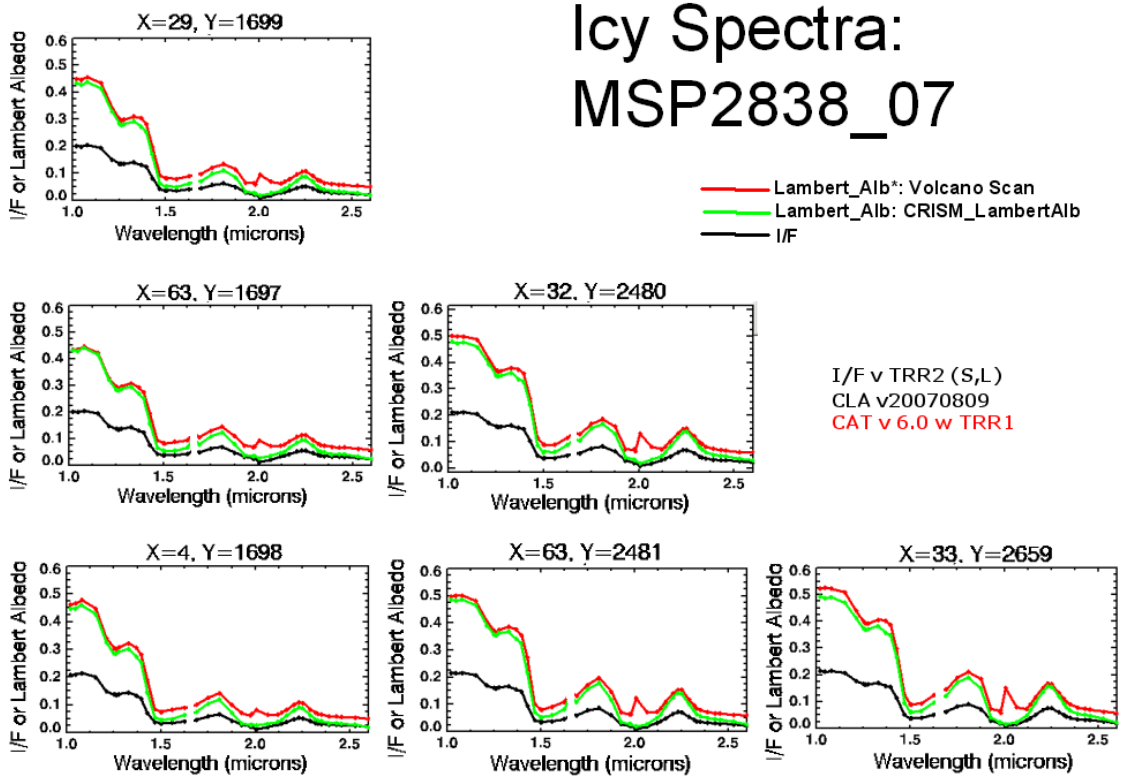

## Figure 8:

We show some of the icy L-detector spectra from the CRISM Multispectral observation, MSP00002838_07, with the same parameters and description as in Figure 7. For clarity, the instrumental artifact near 1.65 μm has been removed from these spectra, though it is somewhat subdued in these icy spectra since the surface is darker at these wavelengths due to the broad $H_2O$-ice absorption band at 1.5 μm. The spectral channels near 2.0 μm are well-corrected by only the DISORT-based technique for these icy spectra, which are darker due to the deep $H_2O$-ice absorption band at 2.0 μm. The offset in Lambert albedo between the two correction techniques over the entire spectral range is caused by aerosol optical depths being explicitly handled by the DISORT-based technique, and not handled by the Volcano-Scan technique. Note that the Lambert albedo spectra as computed by the DISORT-based technique (CRISM_LambertAlb) often appear to be closer to the *I/F* spectra for the darker wavelengths, but these same DISORT-corrected Lambert albedo spectra also often appear closer to the Volcano-Scan corrected Lambert albedo spectra for the brighter wavelengths.

## 5.c) Comparison of Map tiles for Tyrrhena Terra

A primary purpose of performing the correction of the CRISM multispectral *I/F* TRDRs for photometric, atmospheric and thermal effects is to be able to mosaic or overlay multispectral TRDRs from different orbits under different observing conditions. If this correction was not performed prior to overlaying the different TRDR strips, then the varying observing conditions like different aerosol optical depths, different photometric angles, and different surface temperatures would cause the different TRDR strips, when juxtaposed, to appear wildly different. By performing this correction for different observing conditions for a number of different TRDR strips and then map-projecting these strips onto a common grid, we construct a 'map tile'. CRISM has 1964 map tiles that span the surface of Mars.

We show a portion of one map tile for Tyrrhena Terra in Figure 9, both prior to correction and after correction for the different observing conditions in each orbit. The three bands chosen for the false-color image of the map tile are: the blue band at 0.86 µm, to assess aerosol correction; the green band at 1.92 µm, for surface hydration mapping; as well as, the red band at 2.01 µm, for evaluating the quality of the correction in this $CO_2$ gas band. This is a particularly challenging set of wavelengths to use, given the proximity of two of the bands (red and green) to the $CO_2$ gas bands at 2.0 microns, as well as the blue band being at a wavelength (0.86 µm) that is strongly affected by aerosols. For this map tile of Tyrrhena Terra, the Lambert-albedo version has significantly better continuity between the different TRDR strips than does the *I/F* version. Indeed, the *I/F* version of the map tile has a few strips that are rather blue in color, which is caused by the uncorrected aerosols. Furthermore, the *I/F* version of the map tile has a significant amount of optical distortion known as spectral smile in the $CO_2$ gas band near the edges of the many of the TRDR strips. We explicitly correct for this spectral smile in the major $CO_2$ gas bands by computing special DISORT ADR AC LUTs for the off-axis imaging in these gas bands. Hence, this smile-induced distortion in the gas bands is largely absent in the Lambert-albedo version of the map tiles.

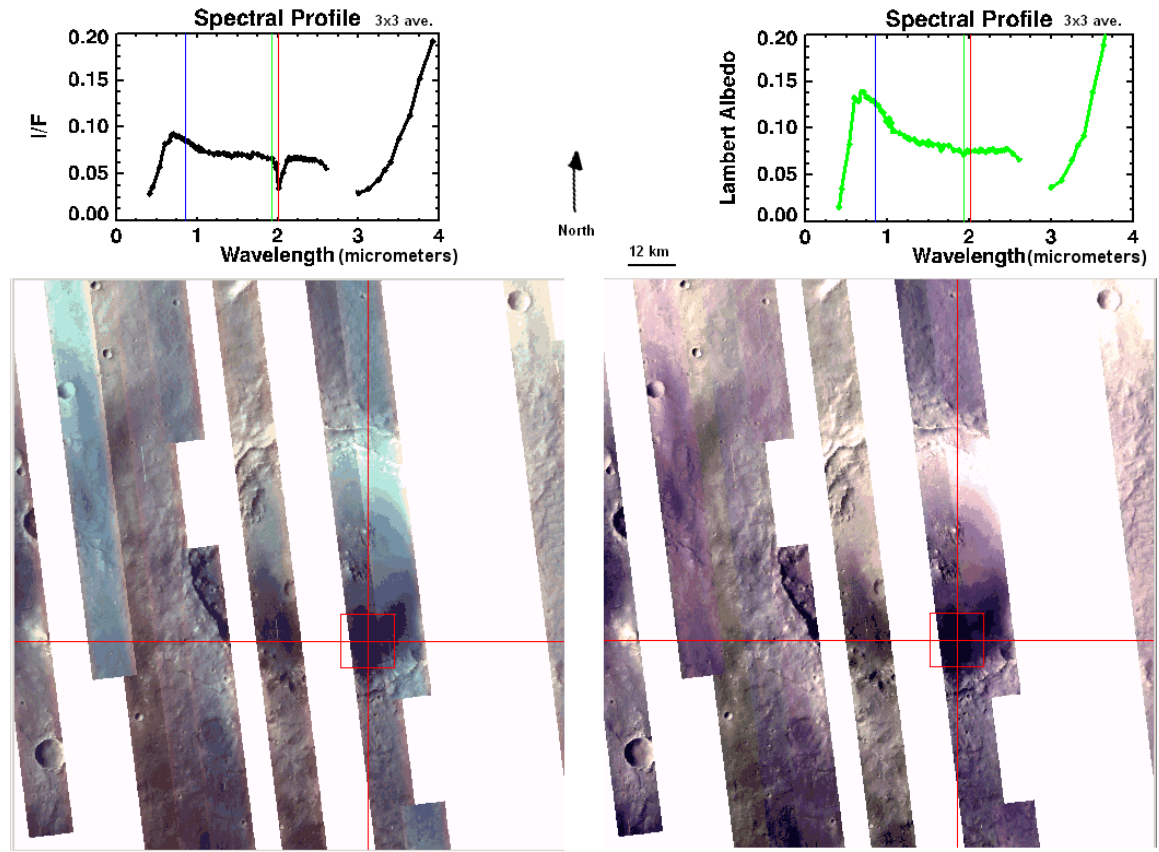

**Figure 9:**

The CRISM multispectral map tile (with both the S & L detectors concatenated), T894, in the Tyrrhena Terra region, northeast of the Hellas basin, is shown, as corrected with the CRISM_LambertAlb DISORT-based software. On the left, we show a portion of the uncorrected *I/F* map tile and the spectrum. The lower-left corner of this image is at 4.50°S, 96.88°E, and the upper-right corner of this image is at 2.50°S, 98.69°E. On the right, we show the corrected version of the same map tile and spectrum, where we used standard pipeline processing and did not correct the photometric angles for surface slope. Thermal correction is not enabled here. North is upwards in both images. Sun direction is somewhat variable, though a westerly incidence angle predominates. The bands shown here are R=2.01 μm, G=1.92 μm, & B=0.86 μm. The stretching of the color space is min-max, with the bounding values chosen independently for each of the R,G,B planes for *I/F* and for Lambert albedo (I/F: R 0.032-0.086, G 0.066-0.155, B 0.073-0.172; Lambert albedo: R=G=B: 0.10-0.30); the bounding values for the color stretching were not chosen to be the same for both *I/F* and Lambert albedo due to the change in range. The spectra shown are 3×3 pixel$^2$ spatial averages centered at the point indicated by the cross-hairs in each map tile. Note that the blue coloring dominates for several of the strips in the *I/F* version, whereas this blue coloring is much reduced after atmospheric correction. Also, note that the reddish coloring is common near the edge of most of the strips in the *I/F* version, whereas this red coloring is much subdued after atmospheric correction. This reddish coloring at the edge of the strips is caused by an optical distortion (spectral smile) in the $CO_2$ gas bands, for which we are explicitly correcting. For clarity, known channels of particularly poor data quality (wavelengths of 0.97-1.05 and 2.66-2.80 μm) have been removed from these spectra.

## 5.d) Map tiles near the Phoenix Lander 2007 landing site

Lastly, in order to show the performance of the CRISM_LambertAlb system in an area of Mars where the atmosphere and surface are somewhat more variable, we have made a map tile of a number of CRISM MSP TRDRs over the regions around the planned Phoenix Lander 2007 landing site. We show a portion of this map tile in Figure 10, both without correction (*I/F*) and with correction ($A_L$). The quality of the correction is visible largely by visual inspection of the mosaicked TRDRs – the *I/F* version for the same image stretch has several TRDR strips that are dark, and several that are bright, whereas for the $A_L$ version, the contrast is much improved, showing at the right-hand-side of the map tile some of the dark spots that are indicative of boulder fields surrounding craters, as first confirmed by the high-resolution HiRISE camera, also on MRO (*Arvidson et al*, 2007). The histograms for the green channel (chosen here to be 1.506 µm) also show that the correction system converts the multimodal histogram[25] for the *I/F* map tile to a unimodal histogram for the $A_L$ map tile, which is further evidence that the correction system is working well. The long, small-amplitude tail for larger albedos than the main peak in the histogram of $A_L$ map tile may be caused by slightly imperfect correction of some of the TRDR strips. *K.Seelos et al.* (2007) explore these map-tile mosaics in more detail.

Similar characterization of Mars Science Laboratory (MSL) landing sites (launch planned for 2009) with CRISM data (using both modes: multispectral mapping and hyperspectral targetd) is being pursued, and already has been quite valuable in helping to narrow down the field of potential MSL sites to those that have interesting minerals, like phyllosilicates. Future landed missions like ExoMars (planned to be launched by the European Space Agency in 2013) will also benefit greatly from mineralogical information afforded by CRISM, especially with atmospheric and thermal correction, as outlined in this paper.

---

[25] The multimodal histogram for the 1.506-µm channel of I/F is roughly a bimodal distribution with additional substructure for each of the main two peaks. The smaller of the main two peaks is caused by one MSP strip that has a roughly 7° lower incidence angle than most of the other strips in the map tile. This 7° lower incidence angle will make the I/F about 20% brighter for a Lambertian surface without an atmosphere. The observed difference in I/F between the lower-incidence angle strip and an overlapping higher-incidence angle strip is about 30%. Therefore, in this case, differences in incidence angle account for about two-thirds of the observed variation in I/F.

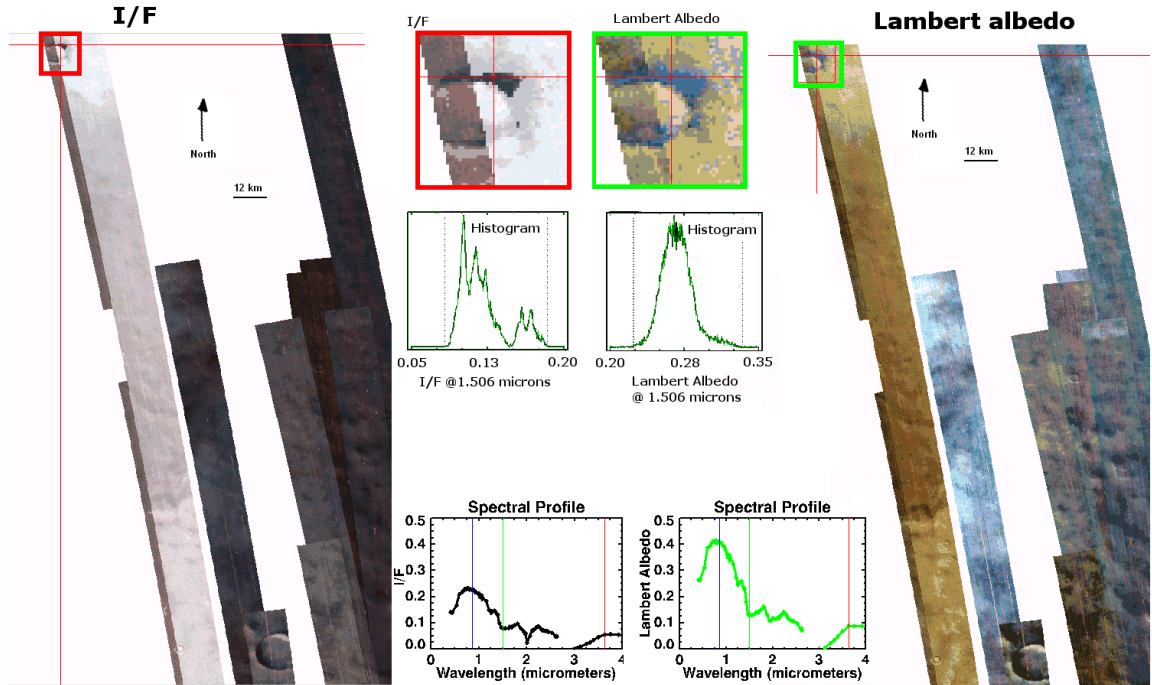

**Figure 10:**
The CRISM multispectral map tile (with both the S & L detectors concatenated), T1889, near the Phoenix landing site region is shown, corrected with the CRISM_LambertAlb DISORT-based software. On the left, we show a portion of the uncorrected *I/F* map tile and the spectrum. The lower-left corner of this image is at 69.21°S, 118.11°W, and the upper-right corner of this image is at 72.50°S, 112.88°W. On the right, we show the corrected version of the same map tile and spectrum. These map tiles were constructed with a summer-only constraint on solar longitude ($L_s$ between 105-165°). North is up in this image. The three color bands in the map tiles are B=0.83 μm, G=1.506 μm, and R=3.6 μm. The stretching of the color space is min-max, with the bounding values chosen independently for each of the R,G,B planes for *I/F* and for Lambert albedo (I/F: R 0.067-0.173, G 0.083-0.184, B 0.092-0.177; Lambert albedo: R 0.187-0.316, G 0.224-0.334, B 0.240-0.364); the bounding values for the color stretching were not chosen to be the same for both *I/F* and Lambert albedo due to the change in range. Each strip in the tiles is approximately 12km in width. In the middle of the figure, we show histograms over the whole map tile of the *I/F* and the Lambert albedo at 1.506 μm. We also show zoom-ins of the ice-rimmed crater at 72.4N, 117.5W, north of the Phoenix landing site; this ice-rimmed crater is the location of the spectra shown here. There is no spatial averaging for these spectra. Note that the portion of the Lambert albedo map tile shown on the right has much better correction for photometric and atmospheric effects than the uncorrected *I/F* version on the left. This is apparent in several ways: (i) much better matching of the strips; (ii) much less structure in the 1.506-μm histograms; and (iii) much better visibility of some of the small dark rock fields in the right side of the Lambert Albedo map tile. For clarity, known channels of poor data quality (wavelengths of 0.97-1.05, 1.65, and 2.66-2.80 μm) have been removed from these spectra.

## *6. Final Remarks & Outlook*

We have summarized in some detail the essential aspects of the system for correcting CRISM multispectral data for variable observing effects such as atmospheric aerosol optical depths, surface pressure, and surface temperature. The uncorrected CRISM *I/F* spectra are used along with the historical climatological observations of the aerosols, pressure and temperature in order to retrieve Lambertian albedo spectra for each CRISM pixel. This system has been applied and tested on innumerable different CRISM multispectral data sets, and four of these tests have been described here. Soon, we will be completing the production of the 1,894 CRISM multispectral map tiles that span the planet.

In Phase II of this work, which is currently being completed, we will add the physical thermal-correction approach (*Martin,* 2004*; Kieffer et al.,* 1977) discussed in section 2. This improvement will allow more accurate correction at smaller spatial scales at wavelengths greater than 3 μm. Also, in Phase II, we will incorporate improved measurements of the aerosol scattering properties, as measured directly by CRISM (*Wolff et al.*, 2007). This improvement will allow more accurate correction at wavelengths shorter than about 1 μm.

After having applied the Phase I + II CRISM_LambertAlb system to the multitudinous multispectral TRDR strips, these map tiles will have much less additional unwanted variability due to shifting observing conditions. Hence, the true spectral diversity and variability can be identified more easily, documented and understood. Indeed, the *I/F* versions of the map tiles are now being used in order to provide more targets for the CRISM hyperspectral targeting mode.

In the phases beyond Phase II, we will consider incorporating:
1. Improved correction in the carbon dioxide gas bands, by directly sensing the carbon dioxide column depth for each pixel, following the techniques of the OMEGA team (*Spiga et al.*, 2007; *Forget et al.*, 2007);
2. Water vapor correction and carbon monoxide correction, either as a part of the pipeline processing, or as a standalone tool;
3. Correction for non-Lambertian effects, for example by using the Lunar-Lambert model (*McEwen et al.*, 1991) or the Hapke model (*Hapke*, 1993);
4. Aerosol correction of the polar regions, by incorporating maps of the polar aerosols, as measured by CRISM emission-phase-function measurements (*Brown, McGuire, & Wolff*, 2008).

As the Lambert albedo versions of the map tiles are completed, spectral ambiguities will become disentangled, which will allow improved CRISM and HiRISE targeting of the martian surface during the Primary Science Phase of the Mars Reconnaissance Orbiter. This disambiguation is especially important near the $CO_2$ gas bands at a wavelength of 2.0 μm, since there are a hydration band at ~1.9 μm and a water ice band at 2.0 μm. The atmospheric aerosol correction will be important for enabling mineralogical retrievals for the shorter wavelengths (shortward of ~1.3 μm), where pyroxene and olivine have different spectral characteristics. Furthermore, the thermal correction will allow the identification of different minerals that have absorptions between 3.0-3.9 μm, including $H_2O$ and $CO_2$ ices, carbonates and organics. Carbonates and organics may not be particularly abundant at the multispectral mapping pixel-scale of 200 meters/pixel, but certainly the ices are abundant on these scales. The multispectral mapping resolution of CRISM will allow us to map these ices and other minerals at unprecedented spatial scales in the visible to near infrared spectral range (0.362-3.92 μm). Hence, the next landed missions to Mars will have ever-increasing prospects for going to the right place for more detailed *in situ* analyses.

**Acknowledgements:** We are grateful for the assistance of Ed Guinness, Tom Stein, Lars Arvidson, Susan Slavney, and Margo Mueller. Much of the work by PCM has been funded by a Robert M. Walker senior research fellowship from the McDonnell Center for the Space Sciences. We all acknowledge support from NASA funds through the Applied Physics Laboratory, under subcontract from the Jet Propulsion Laboratory through JPL Contract #1277793. SMW acknowledges support from a NASA Graduate Student Research Program fellowship. TNT acknowledges support through the MRO Participating Scientist award #1300367, as well as support through the MGS-TES and ODY/THEMIS projects. TZM, ROG and RLM acknowledge support through the MRO project.

## *Appendix: List of Acronyms[26]:*

| | |
|---|---|
| $A_L$ | Lambert albedo |
| ADR | Ancillary Data Record |
| ADR TE | ADR Thermal Emission |
| ADR AC | ADR Atmospheric Correction |
| ADR CL | ADR Climatology |
| CAT | CRISM Analysis Tool |
| CRISM | Compact Reconnaissance Imaging Spectrometer for Mars (which is on-board MRO) |
| CRISM_LambertAlb | Name of the pipeline software system described here |
| DISORT | DIScrete-Ordinate-method Radiative Transfer |
| EDR | Experimental Data Record |
| EMI | Emission or Emergence angle |
| GRS | Gamma Ray Spectrometer (which is on-board Mars Odyssey) |
| HiRISE | High Resolution Imaging Science Experiment (which is on-board MRO) |
| HRSC | High-Resolution Stereo Camera (which is on-board Mars Express) |
| INC | Incidence angle |
| I/F | measured radiance divided by solar irradiance |
| L detector | Long-wavelength detector for CRISM |

---

[26] We also define some of these acronyms in the text, for ease of reading. They are defined here, in order to have a comprehensive list, for easy reference.

| | |
|---|---|
| LUT | Look-up table |
| MARSIS | Mars Advanced Radar for Subsurface and Ionosphere Sounding (which is on-board Mars Express) |
| MGS | Mars Global Surveyor |
| MGS-TES[27] | Mars Global Surveyor – Thermal Emission Spectrometer |
| MOC | Mars Orbiter Camera (which was on-board MGS) |
| MOLA | Mars Orbiter Laser Altimeter (which was on-board Mars Global Surveyor) |
| MRDR | Mapped Reduced Data Record |
| MRO | Mars Reconnaissance Orbiter |
| MSL | Mars Science Laboratory |
| MSP | Multispectral |
| OMEGA | Observatoire pour la Minéralogie, l'Eau, les Glaces et l'Activité (which is on-board Mars Express) |
| PHI | Azimuthal angle |
| PDS | Planetary Data System |
| PFS | Planetary Fourier Spectrometer (which is on-board Mars Express) |
| S detector | Short-wavelength detector for CRISM |
| SHARAD | Shallow Radar sounder (which is on-board MRO) |
| THEMIS | Thermal Emission Imaging System (which was on-board Mars Odyssey) |
| TRDR | Targeted Reduced Data Record |
| VL | Viking Lander |

---

[27] This acronym is used instead of 'TES', in order to distinguish this instrument from instruments of the same acronym on other spacecraft (i.e. the Tropospheric Emission Spectrometer, on-board the Aura satellite, orbiting the Earth).

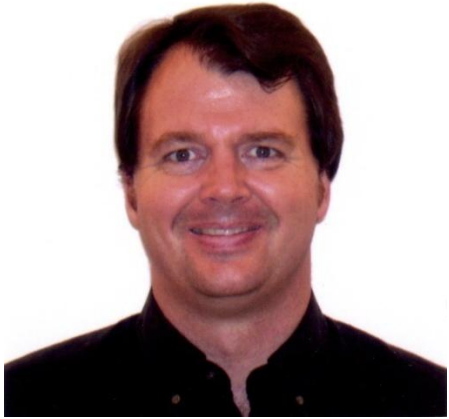

Patrick McGuire received a B.A. in Physics (with Honors) and Mathematics from the University of Chicago and a Ph.D. in Physics from the University of Arizona. Prior to working on the CRISM project, he worked in the fields of computer vision, robotics, neural networks and astrobiology in Bielefeld, Germany, and in Madrid, Spain. His current interests include computer vision, ice spectroscopy and martian polar processes.

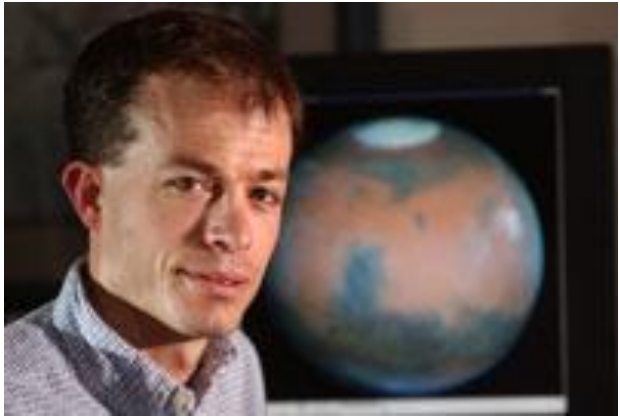

Michael Wolff received a B. S. (summa cum laude) in Physics and Mathematics from the Marquette University and a Ph. D. (magna cum laude) in Astronomy from the University of Wisconsin–Madison. His current interests include radiative transfer and studies of aerosols in the martian atmosphere.

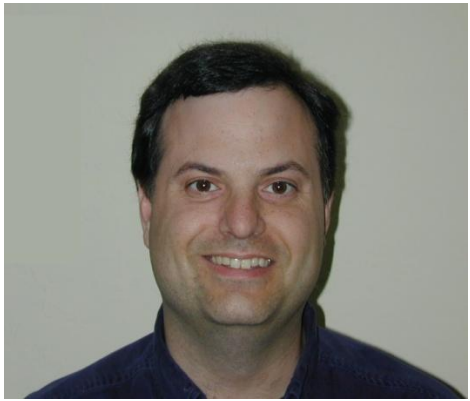

Michael Smith received a B.S. in Physics from the California Institute of Technology and an M.S. and a Ph.D. in Astronomy from Cornell University. His research interests include the meteorology and dynamics of planetary atmospheres, radiative transfer, and remote sensing techniques.

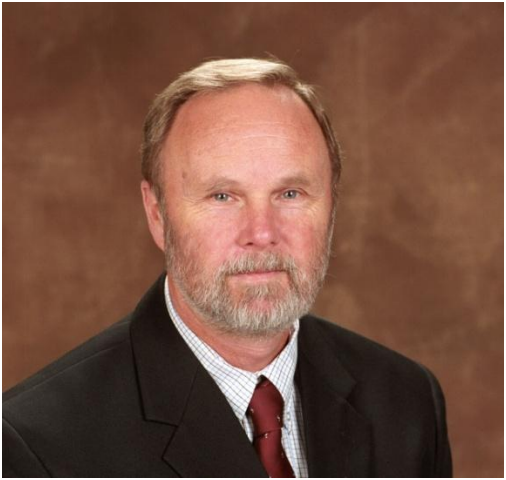Ray Arvidson received an A.B. from Temple University, and an A.M. and a Ph.D. from Brown University, and is a Professor of Earth and Planetary Sciences and a researcher focused on the remote sensing of planetary surfaces.

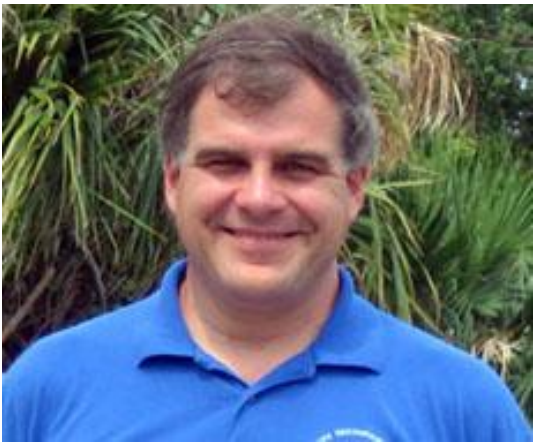Scott Murchie received a B.A. from Colby College, Waterville, Maine, an M.S. from the University of Minnesota, Minneapolis, Minnesota, and a Ph.D. from Brown University, Providence, Rhode Island. He is a planetary scientist whose fields of interest include the evolution of the Martian surface, environments of past Martian water, stratigraphy of terrestrial planet crustal materials, as well as studies of asteroids and small satellites.

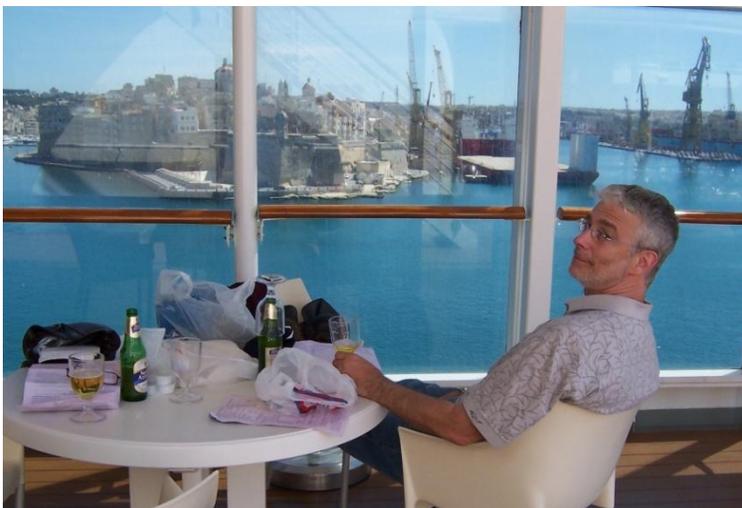Todd Clancy received a B.A. in Geology from the University of North Carolina, an M.S. in Geophysics at Cornell, and a Ph.D. in Planetary Science from California Institute of Technology. His interests include ground-based spectroscopy of planetary atmospheres, photochemistry of the terrestrial atmospheres, and aerosol remote sensing.

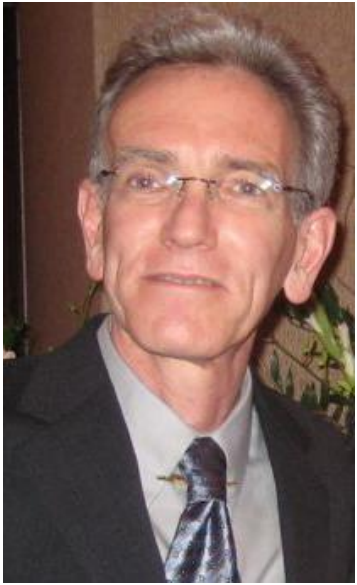

Ted Roush received a B.S. in Geology from the University of Washington and an M.S. and a Ph.D. in Geology and Geophysics from the University of Hawaii. His research interests include using spacecraft and telescopic observations, laboratory simulations, and theoretical calculations, to provide quantitative compositional interpretation of diverse bodies from Mercury to Pluto.

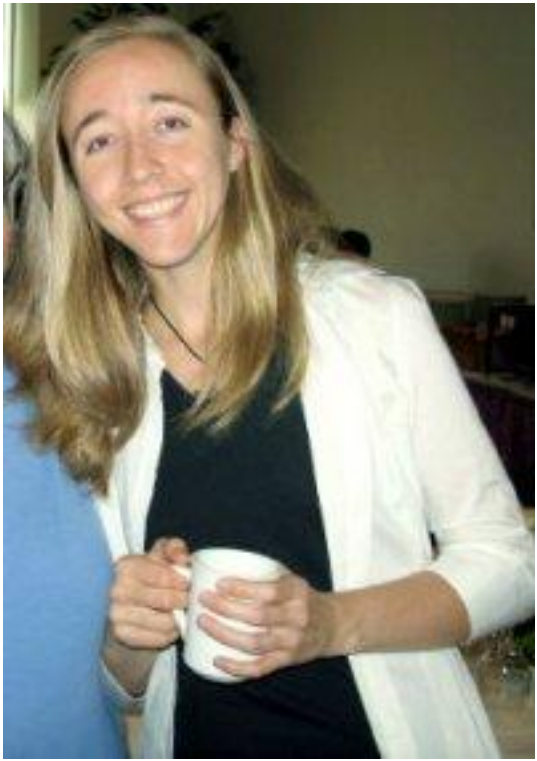

Selby Cull received a B.A. in Geology from Hampshire College, and an M.S. in Science Writing from the Massachusetts Institute of Technology. She is currently a Ph.D. candidate at Washington University in St. Louis.

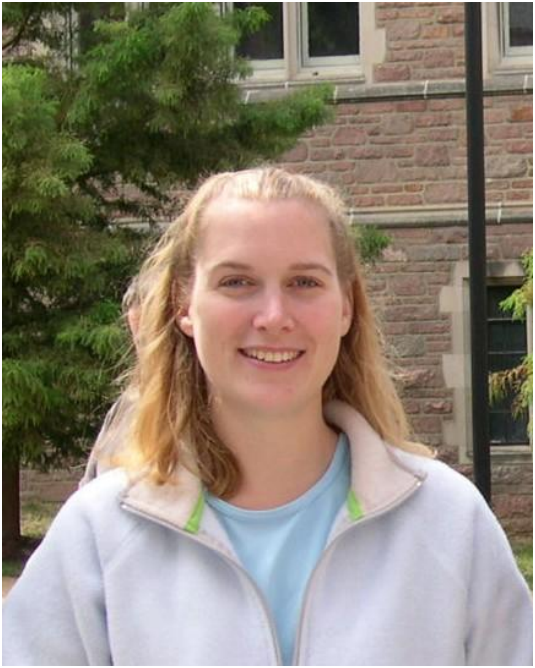

Kim Lichtenberg received a B.S. in Engineering Science from the University of Virginia and an A.M. in Earth and Planetary Sciences from Washington University in St. Louis. She is currently a Ph.D. student and is working on mineralogy and stratigraphy of the martian crust.

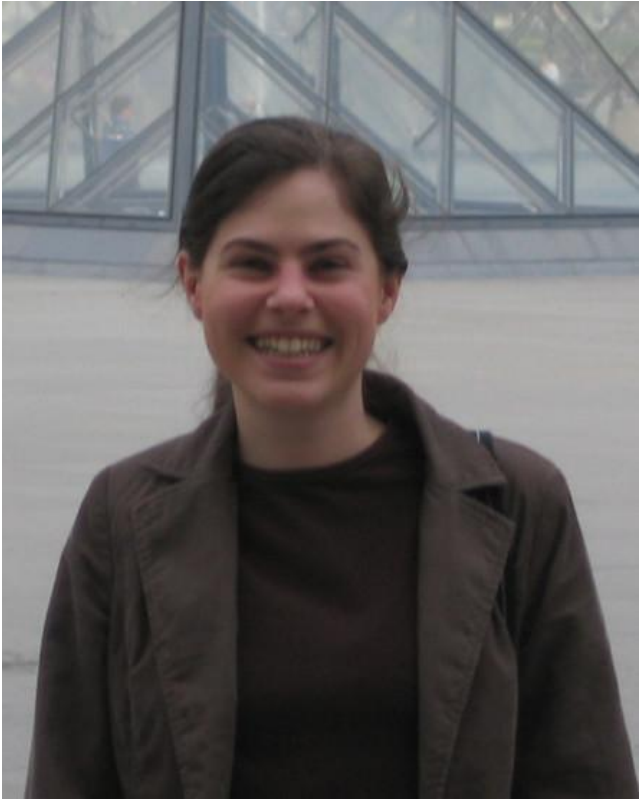

Sandra Wiseman received a B.S. in Geology from University of Tennessee, Knoxville and an A.M. in Earth and Planetary Sciences from Washington University in Saint Louis, and is currently a PhD candidate.

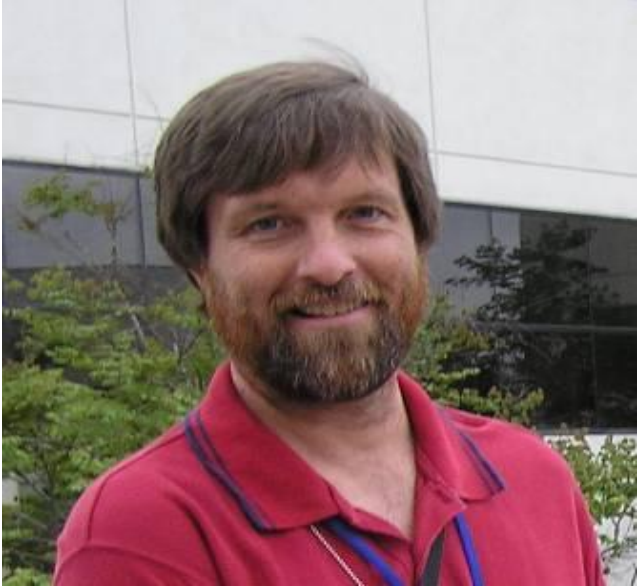

Robert Green received a B.Sc. and an M.Sc. from Stanford University and a Ph.D. from the University of California, Santa Barbara, investigating the spectroscopy of the three phases of water. His research area is environmental imaging spectroscopy with a focus on water and also measurement calibration and validation. As a JPL Senior Research Scientist, he is the experiment scientist for the NASA AVIRIS airborne imaging spectrometer as well as a co-investigator on the CRISM imaging spectrometer for Mars and the M3 imaging spectrometer for the Moon.

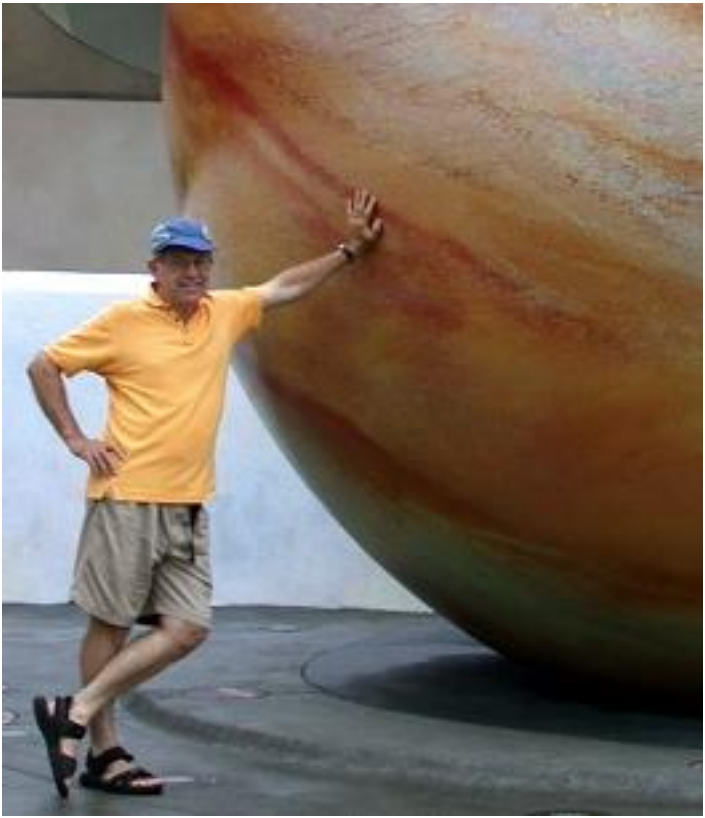

Terry Martin received an A.B. from University of California at Berkeley and a Ph.D. from the University of Hawaii. He is a planetary scientist, specializing in martian surface and atmospheric behavior.

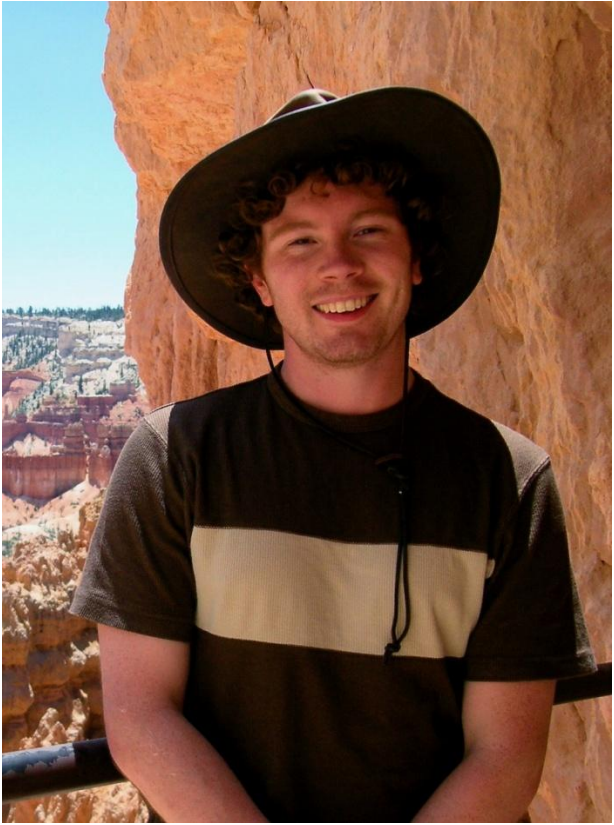

Ralph Milliken received a B.S. from Indiana University and a M.S. and Ph.D. from Brown University in Geology. He is currently a researcher at the Jet Propulsion Laboratory, California Institute of Technology. His research interests include remote sensing, mineralogy, and sedimentology/stratigraphy.

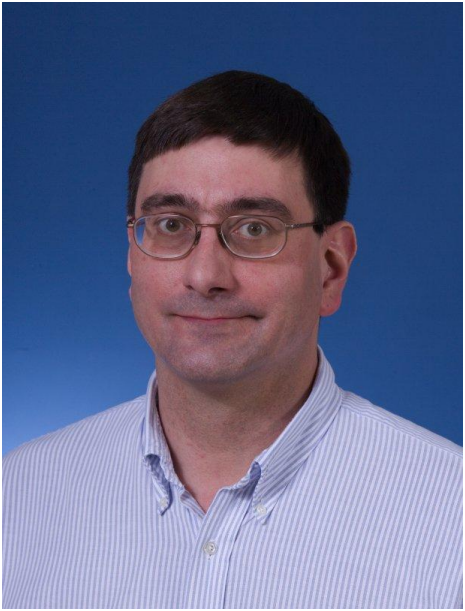

Peter Cavender is a computer engineer who specializes in ground support equipment and signal processing

47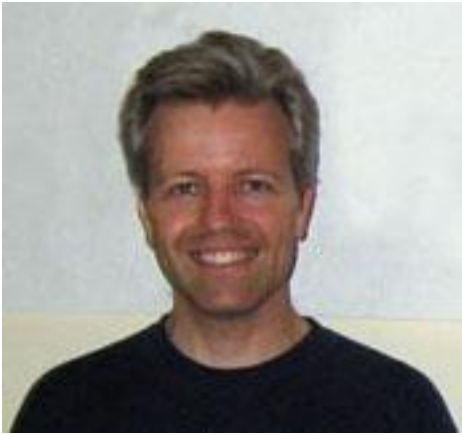

David Humm received a B.S. in both Physics and Astronomy from the University of Iowa and an M.S. and a Ph.D. in Physics from the University of Illinois, Urbana-Champaign. As instrument scientist, he has calibrated imaging spectrometers and other space-based optical instruments from the far ultraviolet to the short-wave infrared.

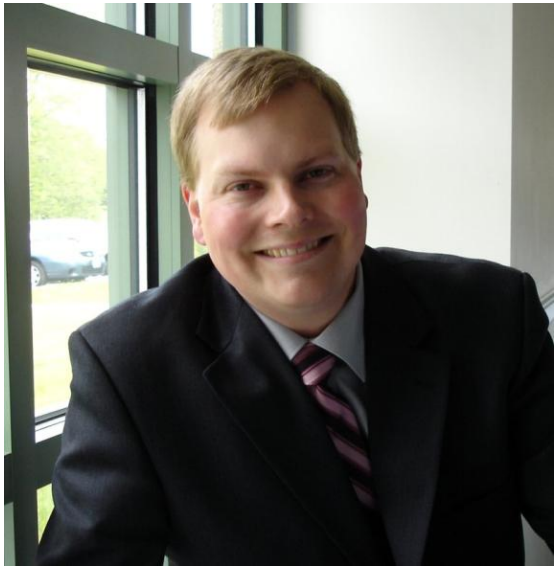

Frank Seelos received a B.S. in Physics and a B.A. in Mathematics from Wofford College, and an A.M. and a Ph.D. in Earth and Planetary Science from Washington University in St. Louis. Frank is the MRO CRISM science operations lead and a planetary scientist at the Johns Hopkins University Applied Physics Laboratory.

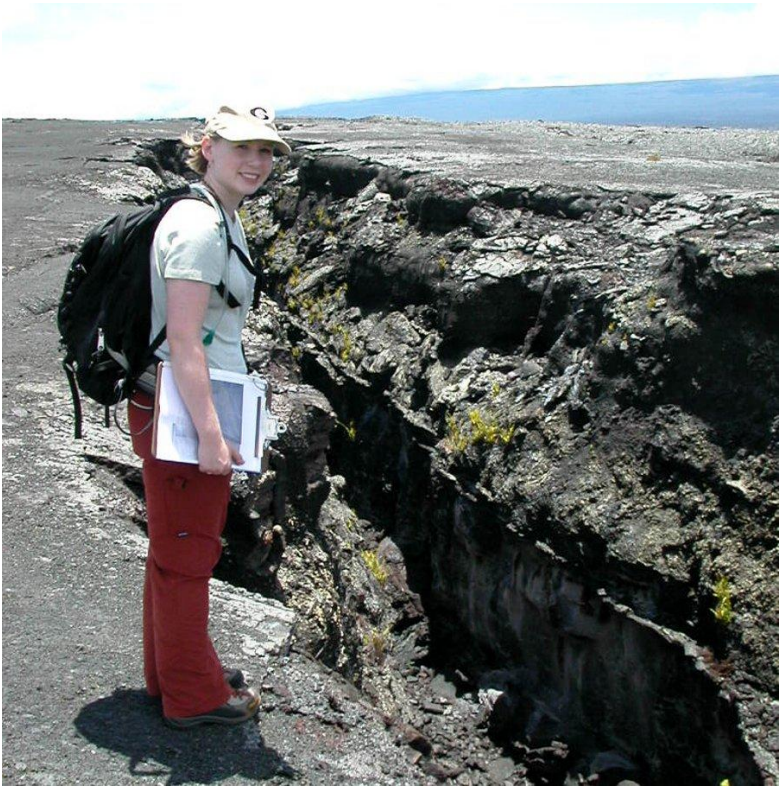

Kim Seelos received a B.S. in Geology from the University of Georgia, and an A.M. and a Ph.D. in Earth and Planetary Science from Washington University in St. Louis, and is a Postdoctoral Research Fellow at the Applied Physics Laboratory of Johns Hopkins University. Kim is involved in both CRISM mission operations and scientific research of the north polar region of Mars.

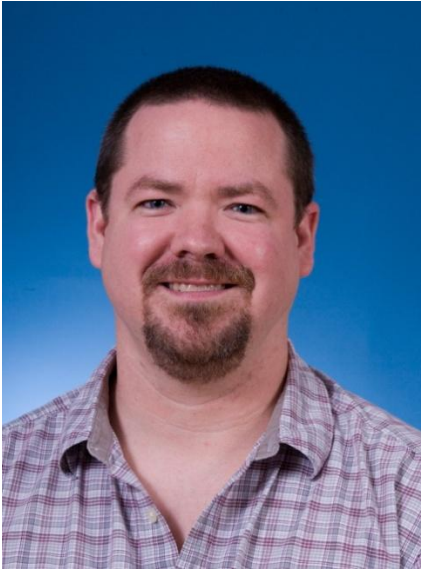

Howard Taylor received a Bachelor and Master of Electrical Engineering from the University of Delaware, specializing in signal and image processing. His interests include data compression, hyperspectral image compression, planetary imager calibration and task automation.

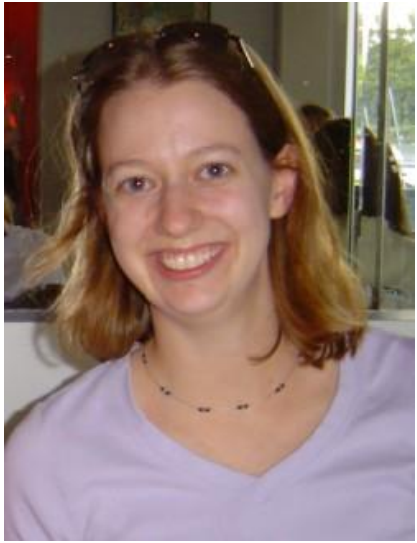

Bethany Ehlmann received an A.B. in Earth and Planetary Science from Washington University in St. Louis and an M.Sc. in Environmental Change and Management and an M.Sc. in Physical Geography, both from the University of Oxford, and is currently a Ph.D. candidate at Brown University.

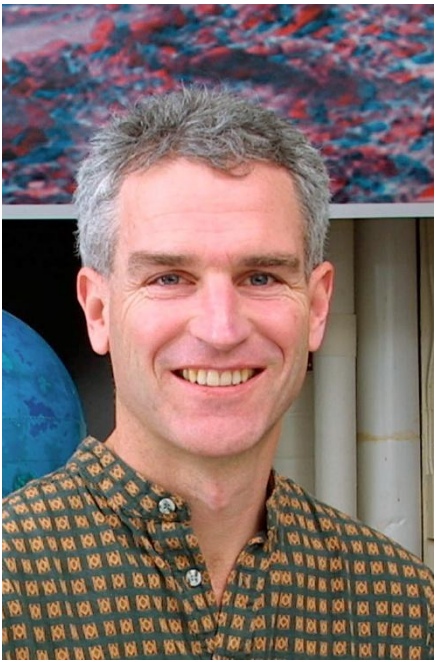

Jack Mustard received a B.Sc. in Geological Sciences from the University of British Columbia, Canada, and an M.Sc. and a Ph.D. in Geological Sciences from Brown University, Providence, Rhode Island. He is currently a Professor of Geological Sciences, a co-investigator with the OMEGA team on the European Mars Express mission and with the CRISM investigation for the NASA Mars Reconnaissance Orbiter.

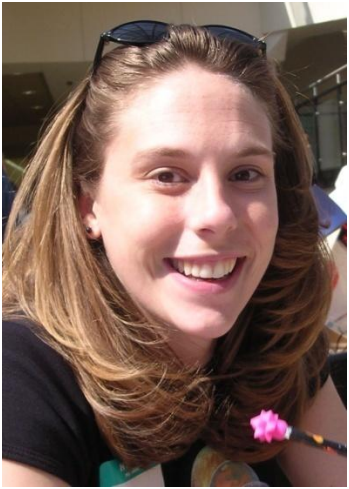
Shannon Pelkey received a B.S. in Astronomy and Physics from University of Massachusetts, Amherst and an M.S. and Ph.D. in Astrophysics and Planetary Science from University of Colorado, Boulder.

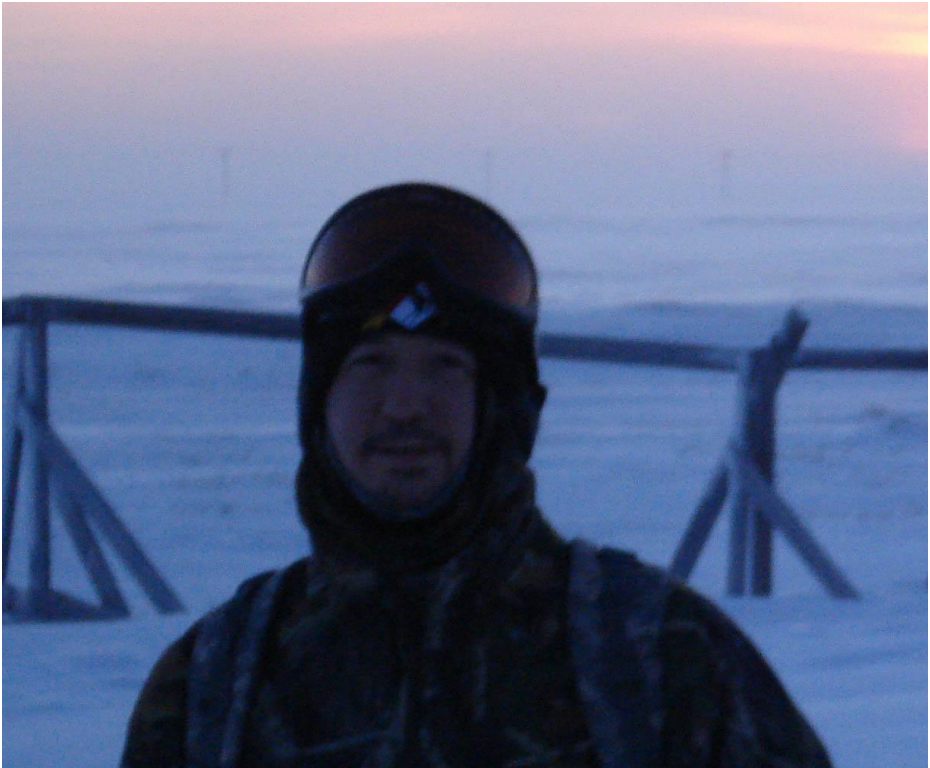
Tim Titus received a B.A. in Physics and Mathematics from Drake University, Des Moines, Iowa, an M.S. in Astrophysics from Iowa State University, Ames, Iowa and a Ph.D. in Astrophysics from the University of Wyoming, Laramie, Wyoming. He is an astrophysicist whose main research interests are martian polar processes, martian aeolian processes, and the development of techniques to detect caves using thermal imagery.

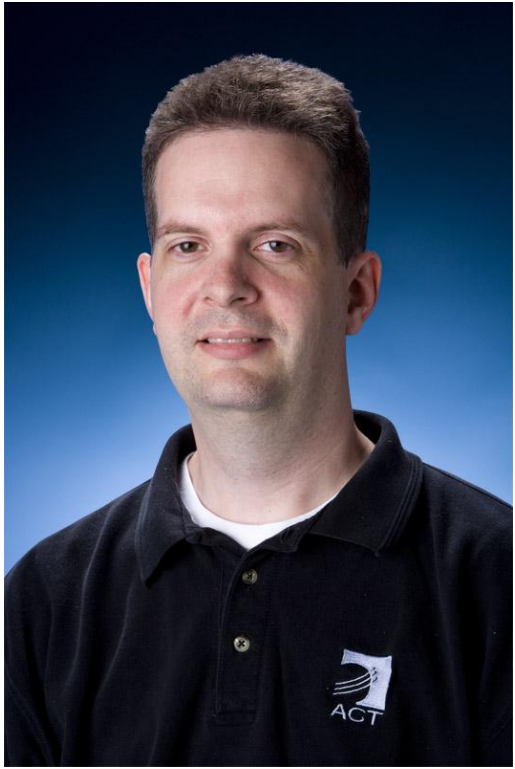

Christopher Hash is a computer scientist specializing in remote sensing data processing software, ground system design and operation. He is the CRISM Science Operations Center downlink operations lead.

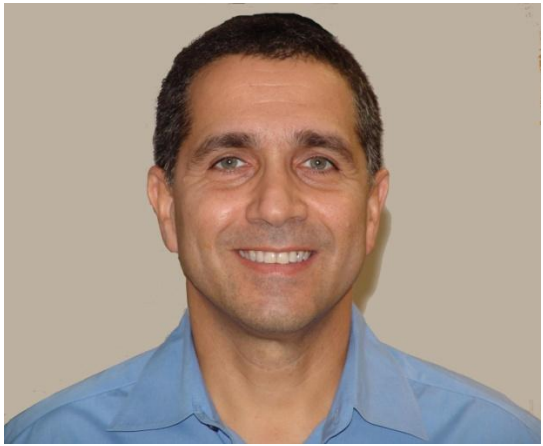

Erick Malaret received a B.Sc in Physics from the University of Puerto Rico, and an M.Sc. and a Ph.D. in Electrical Engineering from Purdue University, Indiana. As the founder and Chief Technology Officer of Applied Coherent Technology Corporation (ACT), he provides software development, algorithm development, system integration, and technical oversight in the area of Image/Signal Processing and Information Technology for aerospace-related programs. He has been (or is) directly involved with following remote sensing related programs: MRO/CRISM, Moon Mineralogy Mapper, MESSENGER, LRO/LROC, Stardust, Deep Impact, MSTI2 & MSTI3, and Clementine. His current interest is Rapid Environmental Assessment.